\documentclass[preprint,12pt]{elsarticle}

\usepackage{lineno,hyperref}
\usepackage{amsmath}
\usepackage{color}
\usepackage{subcaption}
\usepackage{graphicx}
\usepackage{caption}
\usepackage{float}
\usepackage[labelfont=bf]{caption}
\usepackage{subcaption}
\usepackage{epstopdf}
\usepackage{enumitem}
\usepackage[normalem]{ulem}
\usepackage{color}
\usepackage{amssymb}

\newcommand\bV{\boldsymbol V}

\usepackage{booktabs}
\usepackage{multirow}
\usepackage{multicol}
\usepackage{array} 
\usepackage{rotating}

\usepackage{stackengine} 
\stackMath

\definecolor{bleudefrance}{rgb}{0.19, 0.55, 0.91}

\modulolinenumbers[5]

\journal{Journal of Preprint}

\begin{document}

\begin{frontmatter}

\title{Deep Learning combined with singular value decomposition to reconstruct databases  in fluid dynamics}

\author{P. D\'iaz$^1$, A. Corrochano$^{1}$,\\ M. L\'opez-Mart\'in$^{2}$ and S. Le Clainche$^1$}
\address{$^1$ E.T.S.I. Aeron\'autica y del Espacio,
Universidad Polit\'ecnica de Madrid, Spain}
\address{$^2$ E.T.S.I.T., Universidad de Valladolid, Valladolid, Spain }


\begin{abstract}

Fluid Dynamics problems are characterized by being multidimensional and nonlinear. Therefore, experiments and numerical simulations are complex and time-consuming. Motivated by this, the need arises to find new techniques to obtain data in a simpler way and in less time. In this article, we present a novel methodology based on physical principles to reconstruct three-, four- and five-dimensional databases from a strongly sparse sensors as input. The methodology consists of combining Single Value Decomposition (SVD) with neural networks. The neural network used is characterized by a simple architecture based on combining two autoencoders that work in parallel and are joined in the last layer. This new algorithm has been proved with three databases with different dimensions and complexities: in an Atmospheric Boundary Layer (ABL) with a turbulence model and in the flow past a two- and a three-dimensional cylinder. 
Summarizing, this work proposes a new hybrid physics-based machine learning model with a simple, robust and generalizable architecture, which allows reconstructing databases from very few sensors and with a very low computational cost.

\end{abstract}


\end{frontmatter}


\section{Introduction}

Nowadays, the aeronautical industry is closely linked to computational fluid dynamics due to the need to create models that simulate the behavior of the aircraft \cite{zyskowski2003aircraft} to create new designs less pollutant and more energetically efficient. Certain test necessary to obtain high-fidelity databases are difficult to carry out, require a lot of time and are also expensive. For example, one of the most challenging areas of any aircraft design is the prediction of the aerodynamic coefficients present before, during and after stall of the aircraft \cite{ciliberti2017aircraft}. To obtain such coefficients multiple wind tunnel \cite{hall2004overview} and flight tests \cite{heinz2020review} are necessary, which are not always feasible due to the increased cost and the requirement of sophisticated infrastructures. Another example where obtaining data through experimental tests is complicated is in acoustic tests, which are motivated by the need to reduce the noise produced by aircraft on takeoff and landing. This type of testing requires hundreds of microphones to measure noise and thousands of meters of cables \cite{veggeberg2009high}, resulting in a bunch of sparse measurements, sometimes difficult to interpret.

Because of these difficulties, there is a growing interest in mathematical methods to represent as realistically as possible the behavior of aircraft with the least number of test as possible \cite{liu1992numerical}. In this way, experimental data would be combined with predicted or simulated data to correct the design prior to the flight test program \cite{wong2008combined}. In addition to the difficulty of experimental configurations, in many cases, the data to be predicted or simulated have to do with the behavior of fluids around certain bodies. These flows are usually highly-dimensional complex flows (in transient and turbulent regime), and their prediction is particularly difficult, since it requires not only a rich learning model, but also a suitable mapping function that generates an intermediate representation in a low-dimensional feature space (latent space) to reduce its computational needs. 

Numerical simulations and experiments are the two methodologies to study in detail complex flows in fluid dynamics problems. But high-fidelity simulations are expensive and experiments sometimes provide unresolved databases. Therefore, it is complicate to understand the flow physics related to specific problems. Hence, there is a spreading interest in obtaining a method that allows to enhance the resolution of databases, or event to reconstruct two- or three-dimensional flow fields from sparse databases with a low computational cost, which would be an advance for existing problems in fluid mechanics.

There is a growing need to find techniques capable of augmenting databases without the need for experimental setups and heavy simulations \cite{brunton2020machine}. That is why the main objective of this project is to develop a methodology capable of reconstructing databases.

In the line of research on database reconstruction, there are some previous works. For instance, in Ref. \cite{guemes2022super}, a neural network called RAndomly-SEEDed super-resolution GAN (RaSeedGAN) capable of switching from a low resolution image to a high resolution image is presented. This network is tested with turbulent fluid and sea temperature. Also, in Ref. \cite{kim2021unsupervised} an unsupervised GAN neural network is tested and compared with the results of different supervised GAN and Convolutional neural networks. Another type of neural network architecture used to reconstruct flow fields is convolutional neural network \cite{bolton2019applications}\cite{fukami2019super}. In Ref. \cite{erichson2020shallow}, they proposed a shallow decoder (SD) and the input data of the SD are catching selecting random sensors on the surface of a cylinder. 

In line with the above-mentioned studies, this methodology must be able to be applied to different problems regardless of the nature of the data. This implies that the results must be obtained without using the equations that define the fluid behavior and that it is necessary to define generalizable architectures. For this purpose, we have designed a new methodology with neural networks and, keeping the same network architecture, we have tested its robutness and generalization capabilities with different databases. 
The main difference between the method proposed in this paper and the other existing methods lies in the network architecture. The architecture of the neural network has been specifically designed to be simple and efficient, being a design guideline to have a network with as few parameters (weights) as possible to reduce the complexity of the model and the possibility of overfitting, considering the small amount of data available to train the model. Additionally, the architecture must address noisy data and provide representation learning capabilities that are always crucial in any deep learning model \cite{bengio2013representation}. That is the reason for proposing a parallel deep learning architecture based on autoencoders with a final integration layer. The proposed architecture is lightweight and simple and can be trained with a relatively small amount of data with a remarkable ability to generalize under different flow dynamics.

Furthermore, when using machine learning tools there is a computational limit \cite{thompson2020computational} because the computers to be used for programming will have limited temporary memory. This means that the algorithm must have a low computational cost. Therefore, we propose a novel, simple, robust and generalizable architecture to reconstruct databases in fluid dynamics. 

The current work is organized as follows. Section \ref{sec:app} explains the application of this paper in fluid mechanics and Section \ref{sec:reshape} is dedicated to the preliminary data preparation.
In Section \ref{sec:meth}, the methodology and architecture of the neural network are explained. In Section \ref{sec:results}, the most relevant results are shown for different databases. In addition, error tables are included in this section. Finally, section \ref{sec:conclusions} includes the conclusions.

\section{Application to Fluid Dynamics}\label{sec:app}

For the development of this research, data from three different complex fluid fields have been used: air in the atmospheric boundary layer (ABL), two- and three-dimensional flow past a circular cylinder.
The three databases have been generated numerically.
The equations governing fluid dynamics problems are the Navier-Stokes equations. These equations for a viscous, incompressible and Newtonian flow are as follows:

\begin{equation}
\nabla \cdot \overrightarrow{V} = 0
\label{eqn:mass}
\end{equation}

\begin{equation}
\frac{\partial v}{\partial t}+ \left( \overrightarrow{V} \cdot \nabla \right) \overrightarrow{V} = - \nabla p + \frac{1}{Re} \vartriangle \overrightarrow{V}
\label{eqn:mov}
\end{equation}

\ 

where $\overrightarrow{V}$ and $p$ are the velocity vector and the pressure, respectively. $Re$ is the Reynolds number defined as $Re= \rho U L / \mu$ , where $U$ is the characteristic velocity, $L$ is the characteristic lenght, $\rho$ is the density of the fluid and $\mu$ is the dynamic viscosity of the fluid. The equations are non-dimensioned with characteristic length and characteristic time, $L$ and $L/U$.

Additionally, for the atmospheric boundary layer database it is necessary to add to the above two equations the energy conservation equation defined as: 

\begin{equation}
\rho c_p \left[ \frac{\partial T}{\partial t} + \left( \overrightarrow{V} \cdot \nabla \right) T \right] = k \nabla^2 T + \phi
\end{equation}

\ 

where $T$ is the temperature of the fluid, $c_p$ is the specific heat, $k$ is the thermal conductivity coefficient and $\phi$ is the dissipation term.

 \subsection{Athmospheric Boundary Layer (ABL) database}

\begin{figure}[H]
	\centering
	\includegraphics[width = 10 cm]{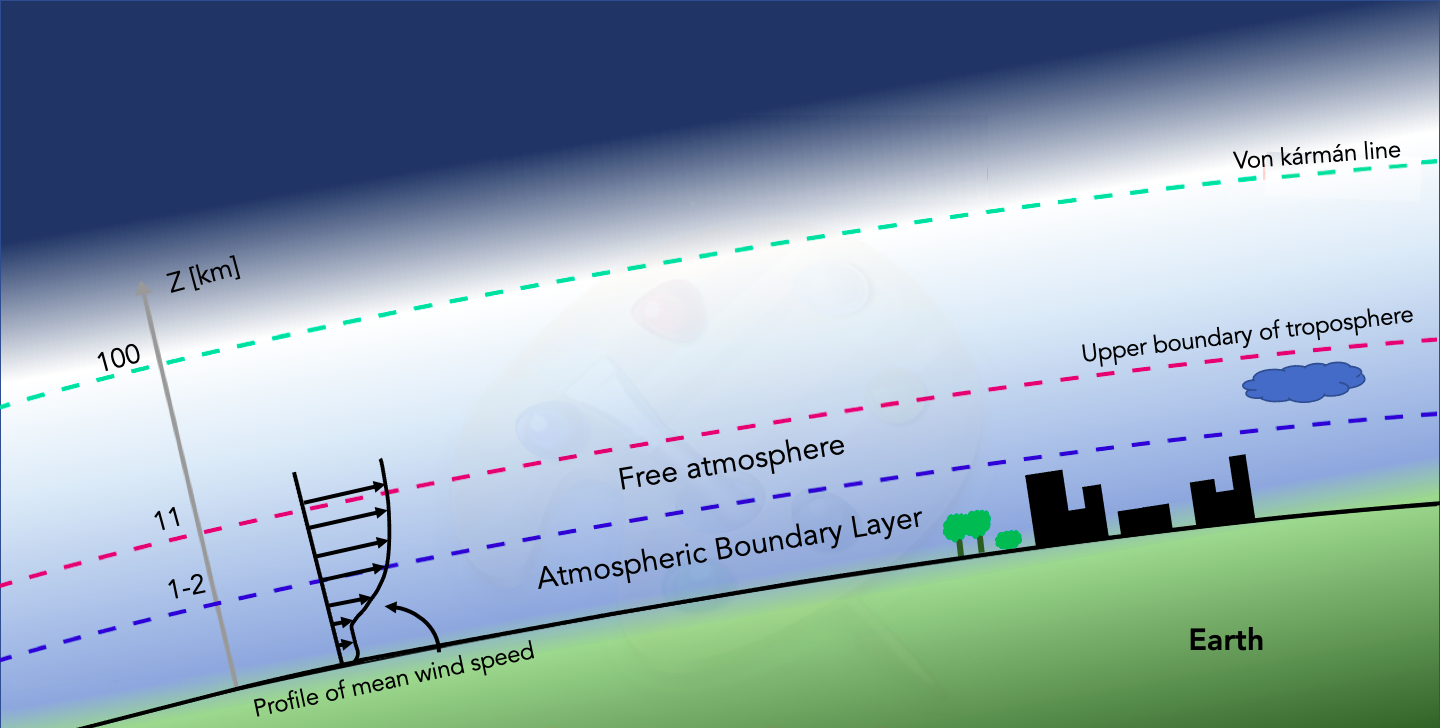}
	\caption{ Sketch of where the atmospheric boundary layer is located in the atmosphere.\label{fig:ABL_layers}}
\end{figure}

A different and relevant problems related to complex flows and with experimental difficulties is the characterization of the air flow in the atmospheric boundary layer (ABL) \cite{garratt1994atmospheric,cuxart2000stable}. This part of the atmosphere is located within the troposphere and occupies approximately 1 km from the earth's surface, see Fig. \ref{fig:ABL_layers}. It is of vital importance to know the characteristics of the air in this area because this is where human life develops, where almost all meteorological phenomena are created and where water vapor, aerosols, pollution, among others, are mixed. Therefore, the flow in the atmospheric boundary layer is highly turbulent and variable, which is why it is challenging to characterize and predict the state variables, such as pressure and temperature, of the air in this area. Understanding the flow behavior can provide great advances in studies related to urban flows, pollutant dispersion and many more applications. 
 
The database is reduced to a small number of points (extracted from the total grid), modelling in this way a set of measurements collected from sensor networks distributed along the terrain. The atmospheric boundary layer is characterized by a turbulent flow that varies according to a large number of variables such as weather conditions, vegetation and buildings in the study area. The variables measured the present study includes  velocity vector, pressure and temperature fields. 

To obtain the database, numerical simulations have been performed. The turbulent flow has been modelled using Reynolds Averaged Navier Stokes (RANS)  k-$\epsilon$ turbulence model, which is commonly used to model the atmospheric boundary layer. The database has been generated using the open source software OpenFoam and has been validated with experiments and simulations. Wall functions have been used to properly model the ABL. More information about the model and the obtained data can be found in Ref. \cite{hargreaves2007use}.

The length of the domain is 5000 m (x-direction) to ensure a developed flow, and the width is limited to 500 m (y-direction), as presented in the literature \cite{hargreaves2007use}. 
The dimension of the ABL database is indicated in Tab. \ref{tab:datadim}, where the three variables correspond to streamwise velocity, pressure and temperature of the air in the atmospheric boundary layer. Furthermore, since the data are averaged in time (RANS model), there is not dimension related to the temporal variable.

\subsection{Circular Cylinder databases}

The analysis of the flow behavior around a circular cylinder is a problem used as a benchmark to validate methodologies. The dynamics of the cylinder is closely linked to the concept of the Re number defined as $Re= \rho U D / \mu$, where $D$ is the diameter of the cylinder and the rest of the variables were defined before. The flow is steady for low Re numbers. From $Re \thicksim 46$, we start to see how a Hofp bifurcation produces an unsteady flow which is lead by a von Karman vortex street \cite{jackson1987finite}. The oscillations remain bidimensional until $Re \thicksim 189$ where a second bifurcation occurs and the flow becomes three-dimensional for some specific wavelengths in the spanwise direction.
For high Re values, the flow becomes turbulent and hence fully three-dimensional \cite{barkley1996three}. 

This behavior is defined by the Navier-Stokes equations described above (Eq. \ref{eqn:mass} and \ref{eqn:mov}). In this research two numerical simulations have been analysed, the two-dimensional cylinder at $Re=100$ and the tree-dimensional cilinder at $Re=280$. In both cases, numerical simulations have been carried out with the domains defined in the literature \cite{barkley1996three} using the open source solver Nek5000 to solve the Navier-Stokes equations. The boundary conditions set in the simulation are Dirichlet for velocity ($v_x = v_y = v_z =0$) and Neumann for pressure. The conditions in the inlet and outlet of the domain are the same: $v_x=1$, $v_y=v_z=0$ and Neumann condition for pressure. The domain of the computational simulations is composed by 600 rectangular elements, each one discretized using the polynomial order $\Pi = 9$. The dimensions of the computational domain  non-dimensionalized with the diameter of the cylinder. The size of the domain in the normal direction is $L_y = ±15D$, and in the streamwise direction are $L_x = 15D$ upstream of the cylinder and $Lx = 50D$ downstream of the cylinder. Details about the database used in this article can be found in Ref. \cite{vega2020higher}.

\begin{itemize}
\item Two-dimensional Cylinder database (Cyl 2D)
    
    This database represents the saturated flow around a 2D cylinder with Reynolds number $Re = 100$ \cite{vega2020higher}. The database analysed is formed by $150$ snapshots, equispaced in time with an interval of $0.2$.
    The dimension of the two-dimensional cylinder database is indicated in Tab. \ref{tab:datadim}. The three variables used in the reconstruction of the database correspond to the two components of the velocity vector $\bV=(u,v)$ (streamwise and normal velocity) and the vorticity $\overrightarrow{\omega} = \overrightarrow{\nabla} \times \overrightarrow{\bV}$. 

\item Three-dimensional Cylinder database (Cyl 3D)
    
    This database represents the value of the velocity of the flow past a circular cylinder at $Re = 280$ \cite{Cyl3D}. 
    A set of 599 snapshots had been collected from the beginning of the numerical simulations equispaced in time with an interval of 1. However, the spanwise velocity does not start to develop until time $350$, and the simulation does not converge (saturated regime) until time $500$ \cite{le2018spatio}. Because of this, only the last $299$ snapshots have been chosen as the representation of the saturated flow regime for the entire database.
    The dimension of the three-dimensional cylinder database is in Tab. \ref{tab:datadim}. Regarding this database, the three variables represent the three components of the velocity (u,v,w). In the three-dimensional flow field, the database has one more dimension than the rest of the databases. 

\end{itemize}

\begin{table}[H]
      \caption{Summary of the dimensions of the databases. $N_{var}$ represents the number of variables. $N_{x}$, $N_{y}$ and $N_{z}$ represent the number of points in x, y and z direction, respectively. $N_t$ represents the number of snapshots. }
        \centering
        \resizebox{6cm}{!} {
        \begin{tabular}{ c | c | c | c | c | c }
         \toprule[0.8mm]
         \textbf{Database} & \textbf{$N_{t}$} & \textbf{$N_y$} & \textbf{$N_x$} & \textbf{$N_z$} & \textbf{$N_{var}$}\\
         \midrule \midrule
         ABL & - & 501 & 2001 & - & 3 \\
         \hline
         Cyl 2D & 151 & 449 & 199 & - & 3\\
         \hline
         Cyl 3D & 299 & 100 & 40 & 64 & 3 \\
         \bottomrule[0.8mm]
        \end{tabular}
     }
     \label{tab:datadim}
\end{table}

\section{Data preparation} \label{sec:reshape}

Before applying the methodology, each database is organized as if the data were collected by a group of sensors (with reduced information). Then, the new methods is applied to reconstruct the original database. The results are compared with the original solution.


The reduction of the databases has been performed in an homogeneous way, i.e., $1$ out of each $k$ points have been chosen in each dimension of the space in such a way that all points are equispaced with respect to each other. Additionally, for the ABL database, the performance of the method has also been tested when only a few points have been chosen in a non-homogeneous way, for instance, concentrated in the wall of the domain (where the strongest changes of the flow occur).

\begin{figure}[H]
	\centering
	\includegraphics[width = 14 cm]{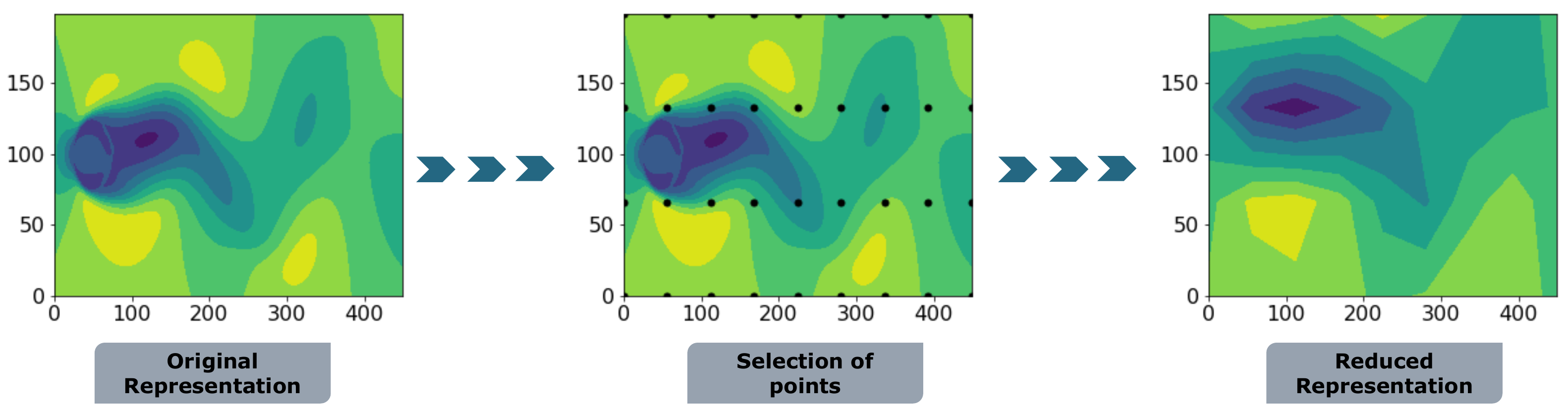}
	\caption{ Example of downsampling to model sensor measurements. Data downsampled equidistant taking 1 every 50 points. Dimension of the original and reduced matrix presentations are $449\times 199$ and $9\times 4$ grid points, respectively. \label{fig:downsampling}}
\end{figure}

In order to better clarify what is done at this step, an example of homogeneous downsampling with the 2D Cylinder database is shown in Fig. \ref{fig:downsampling}. In the first plot we have the original representation of the streamwise velocity. By downsampling, a scale of type 1:k is chosen. In the case of Fig. \ref{fig:downsampling}, a downsampling scale of 1:50 has been chosen, i.e. 1 point out of 50 points is chosen. Since this basis consisted of 449 points in $x$ and 199 points in $y$, the result of reducing the basis by 1:50 gives 9 points in $x$ and 4 points in $y$. The representation of the reduced matrix of $9 \times 4$ is the one shown in the rightmost graph of Fig. \ref{fig:downsampling}.

\begin{figure}[H].
	\centering
	\includegraphics[width = 5 cm]{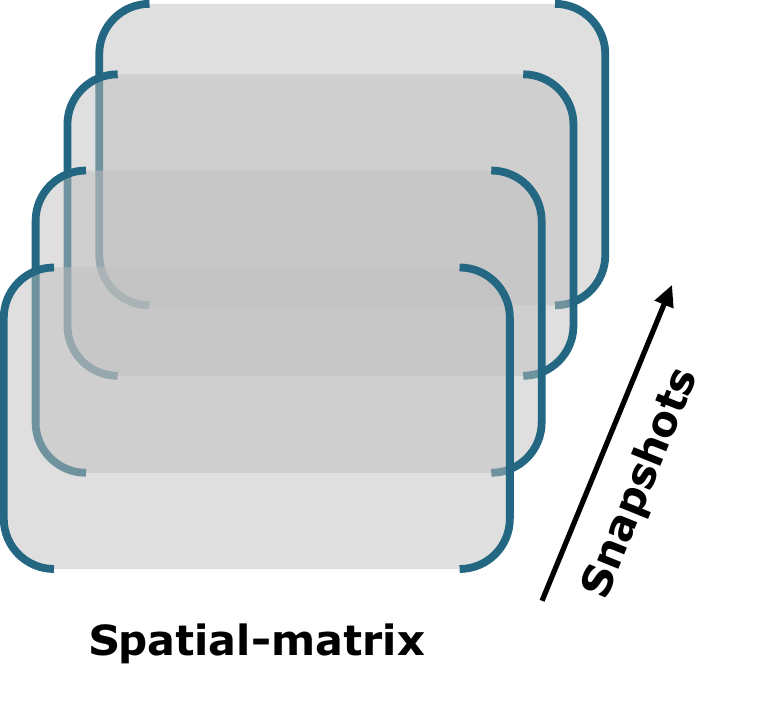}
	\caption{Sketch of matrix organization for one variable. \label{fig:scheme_matrix}}
\end{figure}

Once the reduction of points in the databases has been performed, it is necessary to reorganize the data before applying the methodology.
The spatial dimensions should be collected in a matrix called a spatial matrix. If the database has only x and y axis, the columns of the spatial matrix will correspond to one axis and the rows to another axis. In case the database is three dimensional in space (i.e. it has points on the three axes x,y,z), two axes must be joined by means of a reshaped dimension. Then, there will be a spatial matrix for each time instant which will generate the different snapshots. This structure (shown in \ref{fig:scheme_matrix}) is repeated for each variable. 
In case there is no temporal dimension, as is the case with the ABL database, it is necessary to create a dummy time by repeating the single spatial matrix a number of times to enlarge data dimensionality, which is necessary to properly address the next steps of the algorithm. 

By performing the steps described above, a reduced and reshaped database $X^{DS}$ is obtained. The data in this database are collected in the form of a tensor:
\begin{equation}
X^{DS}=X^{DS}_{ijkl}
\label{eqn:DS}
\end{equation} 
for $i=1,...,N_{t}$, $j=1,...,N_{sy}$, $k=1,...,\hat{N_{sx}}$, $l=1,...,N_{var}$.
Where $\hat{N_{sx}}=N_{sx}$ in the case of ABL and Cyl 2D databases and $\hat{N_{sx}}=N_{sx} \cdot N_z$ in the case Cyl 3D database. \ref{tab:datadim} and $N_{sx}$ and $N_{sy}$ are the numbers of points selected in the downsampling for $x$ and $y$, respectively, which varies for the different test carried out, as it will be specified in the next section.

In the same way, the original tensor of the databases ($X$) is created and then compared with the results of the neural network:
\begin{equation}
X=X_{imnl}
\label{eqn:original}
\end{equation} 
for $i=1,...,N_{t}$, $m=1,...,N_{y}$, $n=1,...,\hat{N_{x}}$, $l=1,...,N_{var}$.
Where $\hat{N_x}=N_x$ for ABL and Cyl 2D databases $\hat{N_x}=N_x \cdot N_z$ for Cyl 3D database. The values of $N_{t}$, $N_y$, $N_{x}$ and $N_{var}$ are indicated in Tab. \ref{tab:datadim}.

\begin{figure}[H].
	\centering
	\includegraphics[width = 14 cm]{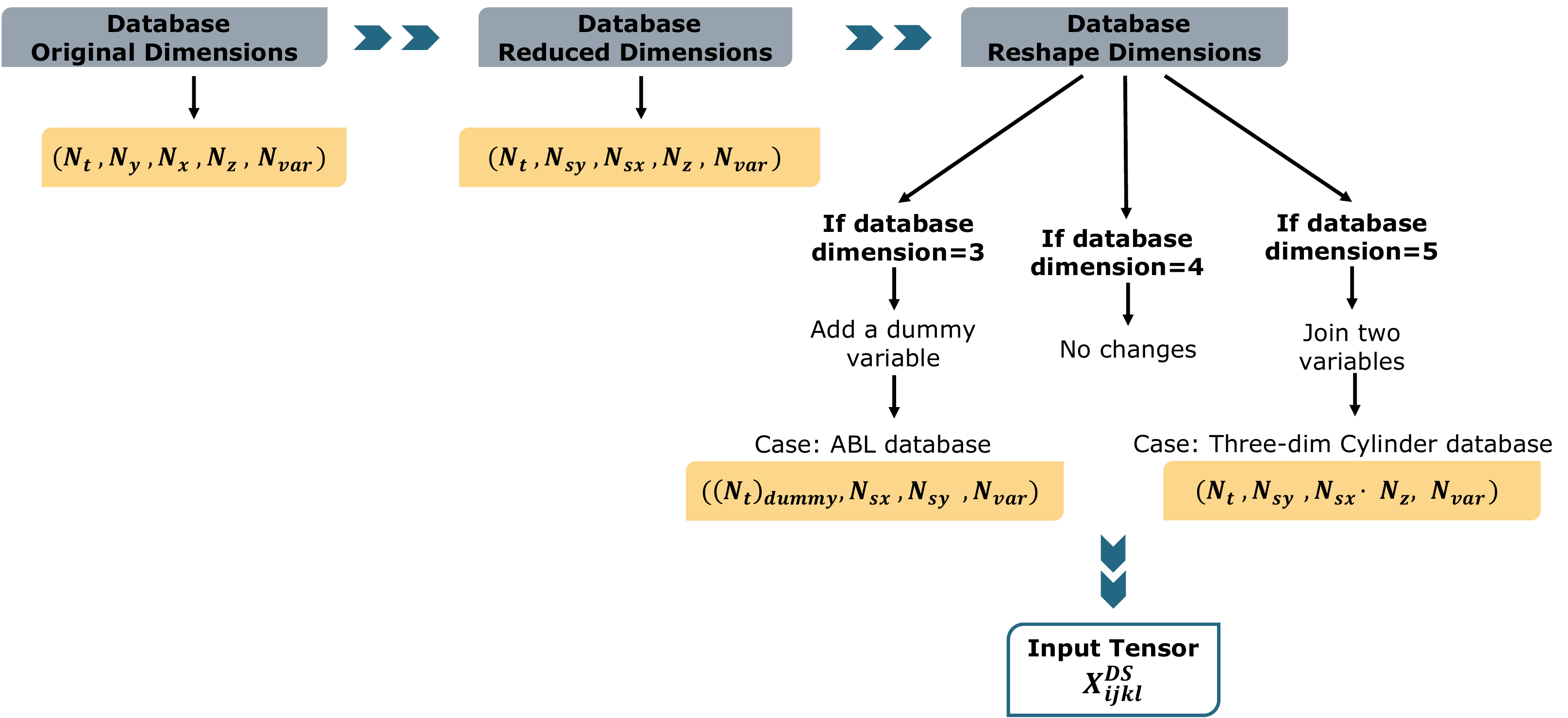}
	\caption{Sketch of the database organization before applying the methodology. The dimensions for each case are indicated in Tab. \ref{tab:datadim}. \label{fig:scheme_org}}
\end{figure}

$X^{DS}$  will be the input to the methodology described in this paper.
Finally, a summary of what has been explained in this section can be seen in the sketch of Fig. \ref{fig:scheme_org}

\section{Methodology \label{sec:meth}}

This section will describe the methodology followed. 
The dimension of a three-, four- or five-dimensional databases has been strongly reduced to a few points, mimicking data collected from experimental sensors. For simplicity, these are called as the reduced databases or reduced tensors ($X^{DS}$). The methodology begins by applying the singular value decomposition method (SVD) \cite{SVDsiro} to the reduced tensor. Subsequently, the result of this decomposition is introduced into the neural network to obtain the reconstructed tensor defined as:

\begin{equation}
\hat{X}=\hat{X}_{imnl}
\label{eqn:reconst}
\end{equation} 

for $i=1,...,N_{t}$, $m=1,...,N_{y}$, $n=1,...,\hat{N_{x}}$, $l=1,...,N_{var}$.

Note that the reconstructed tensor $\hat{X}$ has the same dimensions than the original one previously described (Eq.\ref{eqn:original}).

The sketch represented in Fig. \ref{fig:scheme_meto} shows the general summary of the methodology used.

  \begin{figure}[H]
	\centering
	\includegraphics[width = 12 cm]{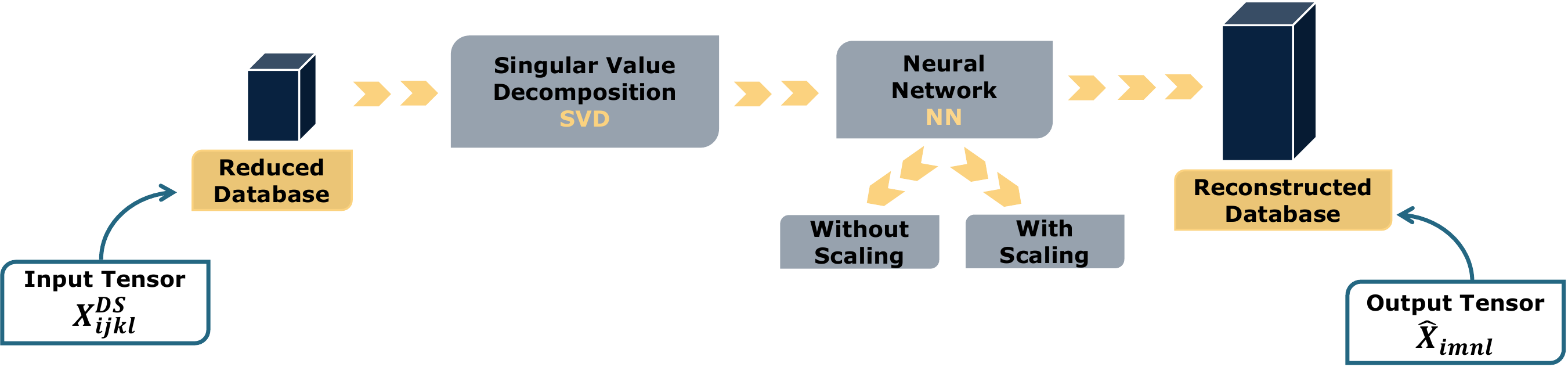}
	\caption{Sketch of the methodology followed. \label{fig:scheme_meto}}
\end{figure}

It should be remembered that the motivation of this study is to obtain an algorithm that allows to increase the number of points in a database which only few points available. Thus, the reduced tensor is the representation of a real database with few points. 
The methodology can be summarized in two main steps, each one corresponding to a part of this section: (i) SVD and  (ii) neural network (NN) architecture. In addition, the performance of the model using a pre-processing techniques has been also evaluated. 

\bigskip

\textbf{Pre-processing techniques}
    
   Before applying SVD, a scaling method is used to pre-process the database, homogenising the range value between all the input variables, avoiding in this way to loose relevant information of the variables with smaller magnitude. Two studies are carried out, scaling and without scaling the database, to test the effect of this pre-processing technique in the final results. Although  we will see that scaling the database does not affect the architecture of the network. 
    
   The scaling method used in the present work ranges each variable between $0$ and $1$, as
    
    \begin{equation}
    X^{SC}=X^{SC}_{l}=\frac{X^{DS}_l-min(X^{DS}_l)}{max(X^{DS}_l)-min(X^{DS}_l)}
    \end{equation}
\smallskip

   for $l=1,...,N_{var}$. Where $min(X^{DS}_l)$ and $max(X^{DS}_l)$ are the minimum and maximum values of the variable $X^{DS}_l$.

\bigskip

\textbf{Accuracy and computational cost}

On one hand, to perform the analysis of the results, two types of characteristic errors will be taken into account: the Mean Square Error (MSE) and the Relative Root Mean Square Error (RRMSE). The MSE is calculated as:

\begin{equation}
MSE=\frac{X-\hat{X}}{N} \label{eqn:MSE}
\end{equation}

\noindent while the RRMSE is computed as

\begin{equation}
	RRMSE=\frac{||X-\hat{X}||}{||X||},\label{eqn:RRMSE}
\end{equation}

\noindent where $X$ is the original database, $\hat{X}$ is the reconstructed database and $N$ is the total number of samples.

On the other hand, the computational cost associated to the model applied is calculated as:
\begin{equation}
t_{c}=t_{1}-t_{0}
\label{eqn:cost}
\end{equation}

  \begin{figure}[H]
	\centering
	\includegraphics[width = 12 cm]{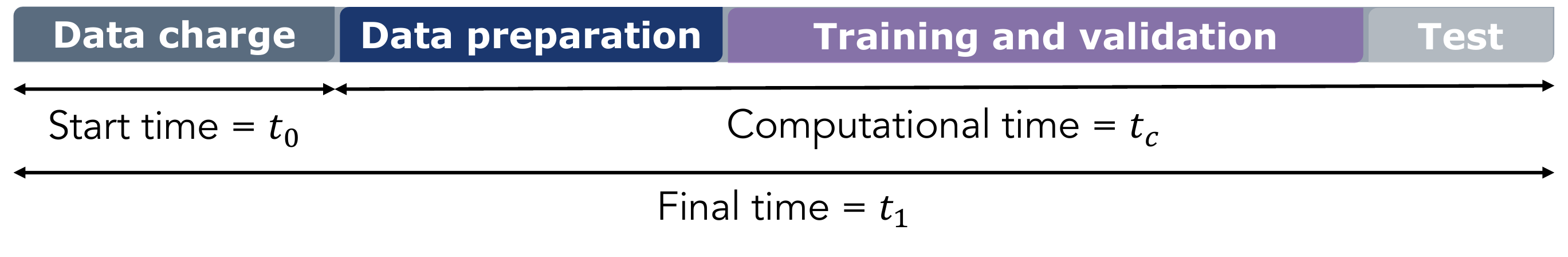}
	\caption{Sketch of the computational cost measurement. \label{fig:scheme_meto}}
\end{figure}

where $t_{1}$ is the final time from the beginning of the code execution to the end of the test part and $t_{0}$ is the initial time from the beginning of the code execution to the end of the data loading.

\subsection{Singular Value Decomposition: SVD}\label{sec:SVD}

Once the reduced (and scaled) tensors are available, the next step in this methodology is to apply a matrix singular value decomposition known as SVD (Single Value Decomposition). This method was developed for fluid dynamics applications by Sirovich \cite{SVDsiro} and consists of a matrix factorization that captures the dominant directions of a database, where the vectors can shrink or grow, depending on the magnitude of the singular values. 
The application of this method is widely extended to various fields of application since it allows to reduce the data dimensionality in complex dynamical systems. The potential of this technique lies in being a method based on simple linear algebra and being easily interpretable. In fluid dynamics, modes identify the main patterns describing the flow physics. Hence, the model presented in this paper considers the SVD modes, so it is based on physical principles.
 
 SVD decomposes a matrix $\Pi$ of dimension $n \times m$ as the product of three matrices: 
 \begin{equation}
 \def\sss{\scriptstyle}
\def\stacktype{L}
 \stackunder{\mathrm{\Pi}}{\sss n\times m}=  \stackunder{U}{\sss n\times n} \  \stackunder{\Sigma}{\sss n\times m} \  \stackunder{V^T}{\sss m\times m}
 \label{eqn:svd}
 \end{equation}
 
 where $()^T$ denotes the matrix transpose and the matrix $\Sigma$ is a diagonal matrix that contains the singular values of the matrix X ordered from greater to smaller magnitude, being all of them greater or equal than 0.

 It is possible to reduce the dimensionality of the data set by compressing the matrix $\Pi$ without losing information, retaining only a few SVD modes, which are associated the largest singular values (modelling the system dynamics).

SVD is applied to each snapshot $i$ of each variable $l$ in such a way that the equation is redefined as follows:

\begin{equation}
X^{DS}_{il}={U}^{il} \ \Sigma^{il} \ (V_{il})^T
\end{equation}

$U$, $\Sigma$ and $V$ matrices resulting from each snapshot $i$ are regrouped in tensor form as follows:

\begin{equation}
X_{U}^{DS}=U_{il}
\end{equation}
 \begin{equation}
X_{S}^{DS}=\Sigma_{il}
\end{equation}
\begin{equation}
X_{V}^{DS}=V_{il}
\end{equation}
\smallskip

for $i=1,...,N_t$ and $l=1,...,N_{var}$. 

The implementation in \textit{Python} of SVD technique has been done by means of the library \textit{numpy.linalg.svd}. This library allows to obtain the $\Sigma$ matrix already reduced with dimensions of $n \times k$ saving computational storage. The $\Sigma$ matrix of dimensions $n \times k$ will be called matrix S. In addition, the $U$ and $V$ matrices would have dimensions of $n \times k$ and $m \times k$ respectively.

\subsection{Neural Network architecture}\label{sec:NNarq}

 Once the SVD method has been applied to the reduced tensors of each database, the next step is to proceed to the reconstruction in order to recover the original dimensions and compare the reconstructed data with the original ones and obtain a percentage error. Different network architectures have been tested, although only the one which provides the best results is described below. 

\begin{figure}[H]
	\centering
	\includegraphics[width = 10 cm]{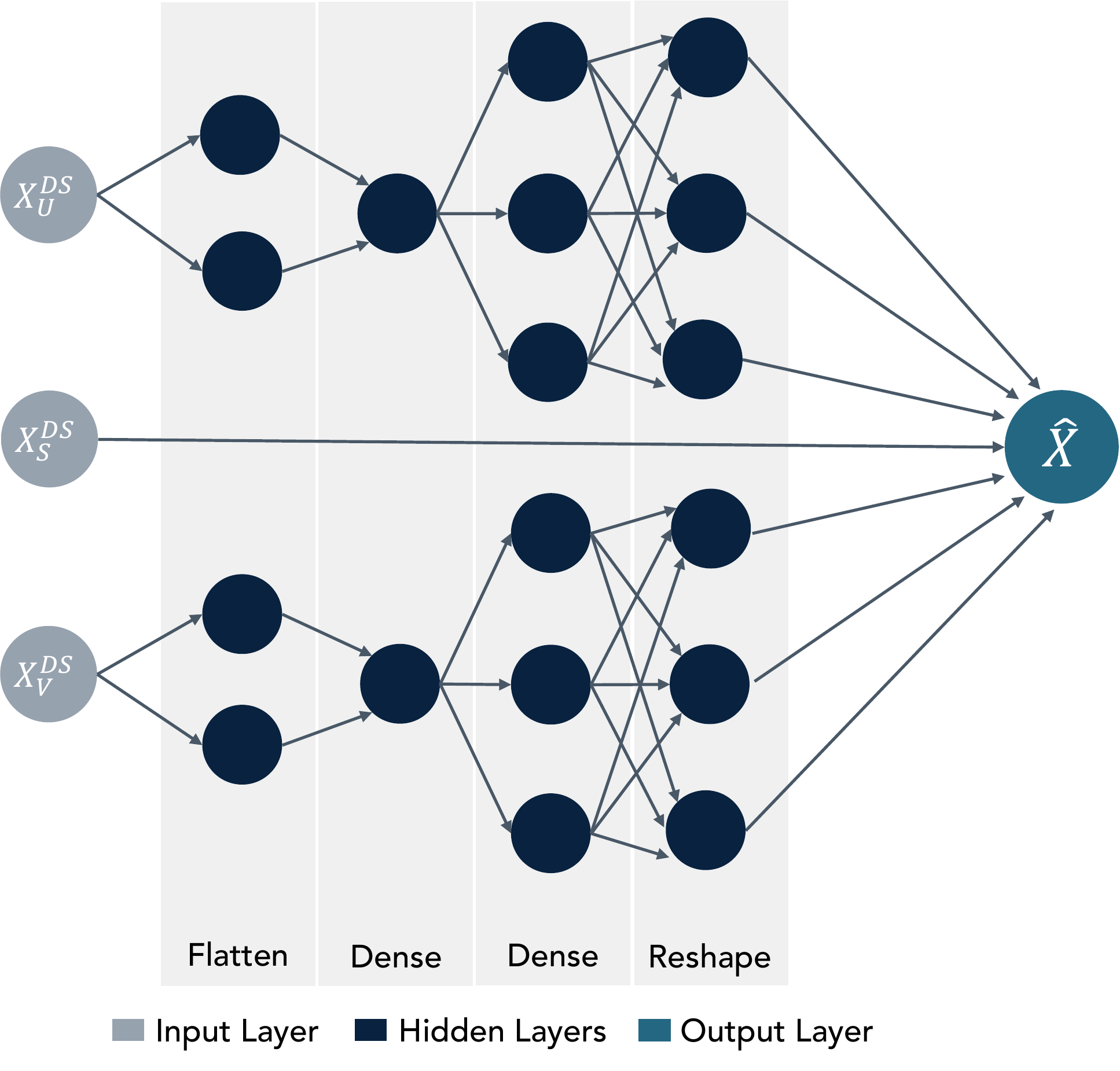}
	\caption{ Sketch of the Neural Network architecture.  \label{fig:NN_arq}}
\end{figure}

 The designed network consists of 3 inputs corresponding to the 3 tensors obtained by the SVD decomposition named $X^{DS}_U$, $X^{DS}_S$ and $X^{DS}_V$, and an output corresponding to the tensor $\hat{X}$ which represents the reconstructed tensor, which has the same dimensions as the original one. Both tensors $X^{DS}_U$ and $X^{DS}_V$ enter an independent autoencoder and on the last layer they are joined, with the tensor $X^{DS}_S$, to form the output tensor. Each autoencoder consists of 5 layers. A sketch of the explained network architecture can be seen in Fig. \ref{fig:NN_arq} and a summary of the layer types is presented in Tab. \ref{tab:NNarq}. In the table there some real variables: $bs$ (batch size), $N_{sy}$ and $\hat{N_{sx}}$ (number of points selected in y-axis and x-axis, respectively), $N_{modes}$ (number of SVD modes) and $N_{var}$ (number of variables).  
 
     \begin{table}[H]
      \caption{Summary of the layers that make up the network used where $bs$ is the batch size; $N_{sy}$ and $\hat{N}_{sx}$ are the number of points selected in y-axis and x-axis, respectively; $N_{modes}$ is the number of SVD modes; $N_{var}$ is the number of variables; $N_{y}$ and $N_{x}$ are the number of points in y- and x- axis in the original databases.}
        \centering
        \resizebox{14cm}{!}{
        \begin{tabular}{ c | c | c | c | c }
         \toprule[0.8mm]
         \textbf{Layer} & \textbf{Layer type} & \textbf{Neurons} & \textbf{Activation function} & \textbf{Dimensions}\\
         \midrule \midrule
         0 & $X^{DS}_{U}$ (Input) &  &  & $(bs,\ N_{sy},\ N_{modes},\ N_{var})$\\
         \hline
         0 & $X^{DS}_{V}$ (Input) &  &  & $(bs,\ N_{modes},\ \hat{N_{sx}},\ N_{var})$\\
         \hline
         0 & $X^{DS}_{S}$ (Input) &  &  & $(bs,\ N_{modes},\ N_{modes},\ N_{var})$\\
         \hline
         1 & u 1 (Flatten) &  &  & $(bs,\ N_{sy} \times N_{modes} \times n_{var})$\\
         \hline
         1 & v 1 (Flatten)&  &  & $(bs,\ N_{modes} \times \hat{N_{sx}} \times N_{var})$\\
         \hline
         2 & u 2 (Dense) & 10 & ReLU & $(bs,\ 10)$\\
         \hline
         2 & v 2 (Dense) & 10 & ReLU & $(bs,\ 10)$\\
         \hline
         3 & u 3 (Dense) & $Ny \times N_{modes} \times N_{var}$ & Linear & $(bs,\ N_y \times N_{modes} \times N_{var})$\\
         \hline
         3 & v 3 (Dense) & $N_{modes} \times N_x \times N_{var}$ & Linear & $(bs,\ N_{modes} \times \hat{N_x} \times N_{var})$\\
         \hline
         4 & $X^{US}_{U}$ (Reshape) & & & $(bs,\ N_y,\ N_{modes},\ N_{var})$\\
         \hline
         4 & $X^{US}_{V}$ (Reshape) & & & $(bs,\ N_{modes},\ \hat{N_x},\ N_{var})$\\
         \hline
         5 & $\hat{X}_{imnl}$ (Out) & & & $(bs,\ N_y,\ \hat{N_x},\ N_{var})$\\
         \bottomrule[0.8mm]
        \end{tabular}
    }
     \label{tab:NNarq}
\end{table}

In any neural network it is necessary to define a series of hyperparameters that define the behavior of the network (Tab. \ref{tab:hiper}). These hyperparameters are the batch size, the loss function and the optimizer.

\begin{table}[H]
\caption{Value of the hyperparameters relevant to different databases.}
\centering
\resizebox{10cm}{!}{
\begin{tabular}{c|cccc}
\toprule[0.8mm]
\textbf{Hyperparameter}  & \multicolumn{1}{c|}{\textbf{ABL}} & \multicolumn{1}{c|}{\textbf{Cyl 2D}} & \multicolumn{2}{c}{\textbf{Cyl 3D}} \\ 
\midrule \midrule
\textbf{Batch size (bs)} & \multicolumn{1}{c|}{3} & \multicolumn{1}{c|}{8}& \multicolumn{2}{c}{16} \\ 
\hline
\textbf{Loss function} & \multicolumn{1}{c|}{MSE} & \multicolumn{1}{c|}{MSE} & \multicolumn{1}{c|}{MSE (Test 1)} & RRMSE (Test 2) \\ 
\hline
\textbf{Optimizer} & \multicolumn{4}{c}{Adam (0.001)} \\ 
\bottomrule[0.8mm]
\end{tabular}
}
\label{tab:hiper}
\end{table}

\bigskip

\begin{figure}[H]
	\centering
	\includegraphics[width = 14 cm]{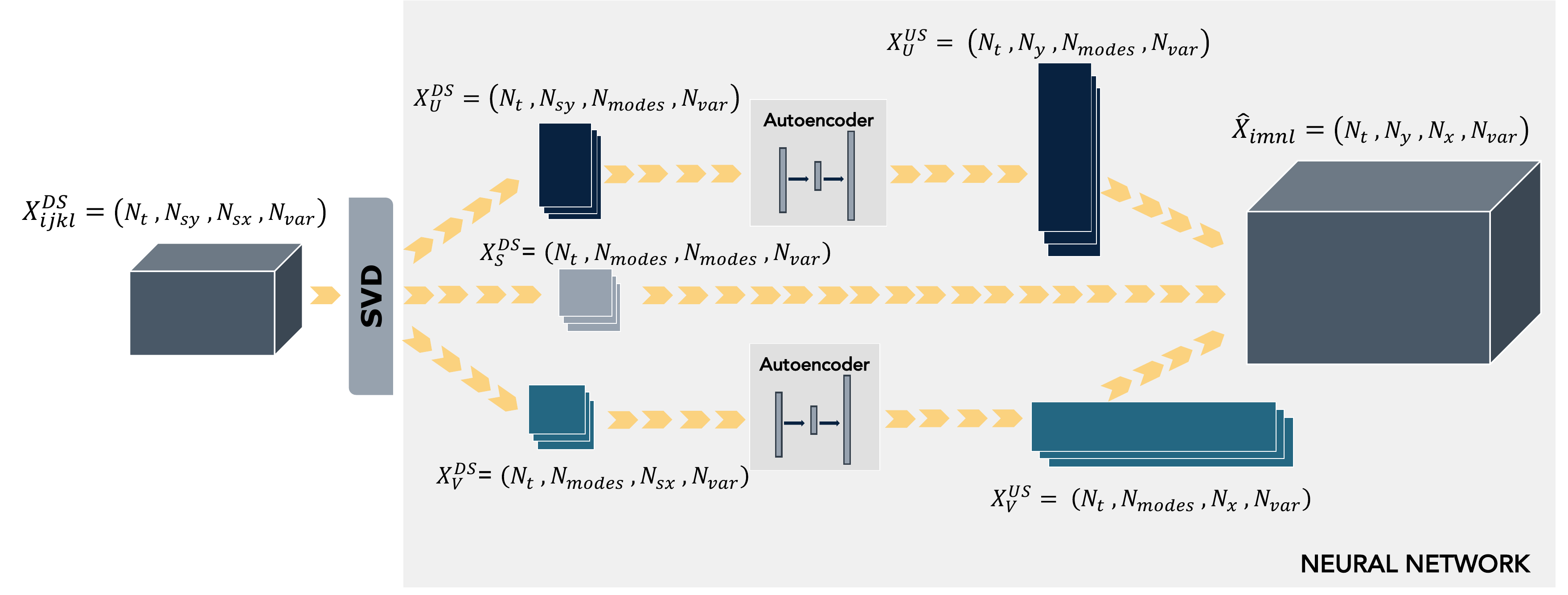}
	\caption{Methodology sketch step by step with dimensions. \label{fig:NNproc}}
\end{figure}

To clarify how the dimensions change along the methodology for different downsamplings $K$ (to set the sensors), Fig. \ref{fig:NNproc} has been included. Depending on which database the methodology is applied to, the dimensions of the tensors involved will change. In Tab. \ref{tab:NNdim}, the dimensions of these tensors have been included for each of the cases included in the results section.
It is necessary to mention that in the case of ABL database the dimension of value 100 corresponds to the number of dummy snapshots ($(N_t)_{dummy}$) and in the case of Cyl 3D database, the dimension corresponding to $N_{x}$ and $N_{sx}$ is equivalent to multiplying these by the dimension $N_{z}$.

\begin{table}[H]
\caption{Summary of the dimensions of the cases show in the results section.}
\begin{tabular}{c|c|c|l|l}
\toprule[0.8mm]
\textbf{Database} & \textbf{k} &\textbf{Tensor} & \multicolumn{2}{c}{\textbf{Tensor dimensions}} \\ 
\midrule \midrule
\multirow{14}{*}{\textbf{ABL}}& \multirow{7}{*}{50}& $X^{DS}$& ($N_t$, $N_{sy}$, $\hat{N}_{sx}$, $N_{var}$) & (100, 11, 41, 3) \\ 
\cline{3-5} 
& & $X^{DS}_{U}$ & ($N_t$, $N_{sy}$, $N_{modes}$, $N_{var}$) & (100, 11, 11, 3) \\ 
\cline{3-5} 
&  & $X^{DS}_{S}$ & ($N_t$, $N_{modes}$, $N_{modes}$, $N_{var}$) &(100, 11, 11, 3) \\ 
\cline{3-5} 
& & $X^{DS}_{V}$ & ($N_t$, $N_{modes}$, $\hat{N}_{sx}$, $N_{var}$) & (100, 41, 11, 3)  \\ 
\cline{3-5} 
& & $X^{US}_{U}$ & ($N_t$, $N_{y}$, $N_{modes}$, $N_{var}$) &(100, 501, 11, 3) \\ 
\cline{3-5} 
& & $X^{US}_{V}$ & ($N_t$, $N_{modes}$, $\hat{N}_{x}$, $N_{var}$) &(100, 11, 2001, 3) \\ 
\cline{3-5} 
& & $\hat{X}$ & ($N_t$, $N_{y}$, $\hat{N}_{x}$, $N_{var}$) & (100, 501, 2001, 3) \\ 
\cline{2-5}
 & \multirow{7}{*}{\begin{tabular}[c]{@{}c@{}} Below: 1:20\\ Above: 1:50\end{tabular}} & $X^{DS}$ & ($N_t$, $N_{sy}$, $\hat{N}_{sx}$, $N_{var}$) & (100, 19, 41, 3) \\ 
\cline{3-5} 
& & $X^{DS}_{U}$ & ($N_t$, $N_{sy}$, $N_{modes}$, $N_{var}$) &(100, 19, 19, 3) \\ 
\cline{3-5} 
& & $X^{DS}_{S}$ & ($N_t$, $N_{modes}$, $N_{modes}$, $N_{var}$) &(100, 19, 19, 3) \\ 
\cline{3-5} 
& & $X^{DS}_{V}$ & ($N_t$, $N_{modes}$, $\hat{N}_{sx}$, $N_{var}$) & (100, 41, 19, 3) \\ 
\cline{3-5} 
& & $X^{US}_{U}$ & ($N_t$, $N_{y}$, $N_{modes}$, $N_{var}$) & (100, 19, 501, 3) \\ 
\cline{3-5} 
& & $X^{US}_{V}$ & ($N_t$, $N_{modes}$, $\hat{N}_{x}$, $N_{var}$) &(100, 2001, 19, 3) \\ 
\cline{3-5} 
& & $\hat{X}$ & ($N_t$, $N_{y}$, $\hat{N}_{x}$, $N_{var}$) &(100, 501, 2001, 3) \\ 
\hline
\multirow{7}{*}{\textbf{Cyl2D}} & \multirow{7}{*}{50} & $X^{DS}$ & ($N_t$, $N_{sy}$, $\hat{N}_{sx}$, $N_{var}$) & (151, 9, 4, 3) \\ 
\cline{3-5} 
& & $X^{DS}_{U}$ & ($N_t$, $N_{sy}$, $N_{modes}$, $N_{var}$) &(151, 9, 4, 3) \\ 
\cline{3-5} 
& & $X^{DS}_{S}$ & ($N_t$, $N_{modes}$, $N_{modes}$, $N_{var}$) &(151, 4, 4, 3) \\ 
\cline{3-5} 
& & $X^{DS}_{V}$ & ($N_t$, $N_{modes}$, $\hat{N}_{sx}$, $N_{var}$) &(151, 4, 4, 3) \\ 
\cline{3-5} 
& & $X^{US}_{U}$ & ($N_t$, $N_{y}$, $N_{modes}$, $N_{var}$) &(151, 449, 4, 3) \\ 
\cline{3-5} 
& & $X^{US}_{V}$ & ($N_t$, $N_{modes}$, $\hat{N}_{x}$, $N_{var}$) &(151, 4, 199, 3) \\ 
\cline{3-5} 
& & $\hat{X}$ & ($N_t$, $N_{y}$, $\hat{N}_{x}$, $N_{var}$) &(151, 449, 199, 3) \\ 
\hline
\multirow{14}{*}{\textbf{Cyl3D}} & \multirow{7}{*}{20} & $X^{DS}$ & ($N_t$, $N_{sy}$, $\hat{N}_{sx}$, $N_{var}$) & (299, 5, 2x64, 3) \\ 
\cline{3-5} 
& & $X^{DS}_{U}$ & ($N_t$, $N_{sy}$, $N_{modes}$, $N_{var}$) &(299, 5, 5, 3) \\ 
\cline{3-5}  
& & $X^{DS}_{S}$ & ($N_t$, $N_{modes}$, $N_{modes}$, $N_{var}$) &(299, 5, 5, 3) \\ 
\cline{3-5} 
& & $X^{DS}_{V}$ & ($N_t$, $N_{modes}$, $\hat{N}_{sx}$, $N_{var}$) &(299, 5, 2x64, 3) \\ 
\cline{3-5} 
& & $X^{US}_{U}$ & ($N_t$, $N_{y}$, $N_{modes}$, $N_{var}$) &(299, 100, 5, 3) \\ 
\cline{3-5} 
& & $X^{US}_{V}$ & ($N_t$, $N_{modes}$, $\hat{N}_{x}$, $N_{var}$) & (299, 5, 40x64, 3) \\ 
\cline{3-5} 
& & $\hat{X}$ & ($N_t$, $N_{y}$, $\hat{N}_{x}$, $N_{var}$) & (299, 100, 40x64, 3) \\ 
\cline{2-5}  
& \multirow{7}{*}{15} & $X^{DS}$ & ($N_t$, $N_{sy}$, $\hat{N}_{sx}$, $N_{var}$)  & (299, 5, 2x64, 3) \\ 
\cline{3-5} 
& & $X^{DS}_{U}$ & ($N_t$, $N_{sy}$, $N_{modes}$, $N_{var}$) &(299, 5, 5, 3) \\ 
\cline{3-5} 
& & $X^{DS}_{S}$ & ($N_t$, $N_{modes}$, $N_{modes}$, $N_{var}$) &(299, 5, 5, 3) \\ 
\cline{3-5} 
& & $X^{DS}_{V}$ & ($N_t$, $N_{modes}$, $\hat{N}_{sx}$, $N_{var}$) &(299, 5, 2x64, 3) \\ 
\cline{3-5} 
& & $X^{US}_{U}$ & ($N_t$, $N_{y}$, $N_{modes}$, $N_{var}$) &(299, 100, 5, 3) \\ 
\cline{3-5} 
& & $X^{US}_{V}$  & ($N_t$, $N_{modes}$, $\hat{N}_{x}$, $N_{var}$) &(299, 5, 40x64, 3) \\ 
\cline{3-5} 
& & $\hat{X}$ & ($N_t$, $N_{y}$, $\hat{N}_{x}$, $N_{var}$) & (299, 100, 40x64, 3) \\ 
\bottomrule[0.8mm]

\end{tabular}
  \label{tab:NNdim}
\end{table}

It is important to understand that the network consists of two autoencoders that work in parallel and meet at the output layer with the input $X^{DS}_{S}$, which only appears at this last part of the neural network. Therefore, the reconstruction error used to improve the weights is calculated by comparing the output tensor $\hat{X}$ with the original tensor, i.e., it is not calculated by comparing the tensors $X^{US}_{U}$ and $X^{US}_{V}$ separately with the original ones. 

The strategy followed to find out if the network is able to predict in time consists of the following. The total number of available snapshots has been divided into three groups: training, validation and test. The snapshots corresponding to the training group will be used to train the network, the snapshots from the validation set will be used to validate the network results into the training part and the snapshots in the test group will be used to test the network after being trained. The distribution of snapshots into the three groups is different to each database. Values for $T_{train}$, $T_{val}$ and $T_{test}$ are giving in Tab. \ref{tab:snap}.

\begin{figure}[H]
	\centering
	\includegraphics[width = 7 cm]{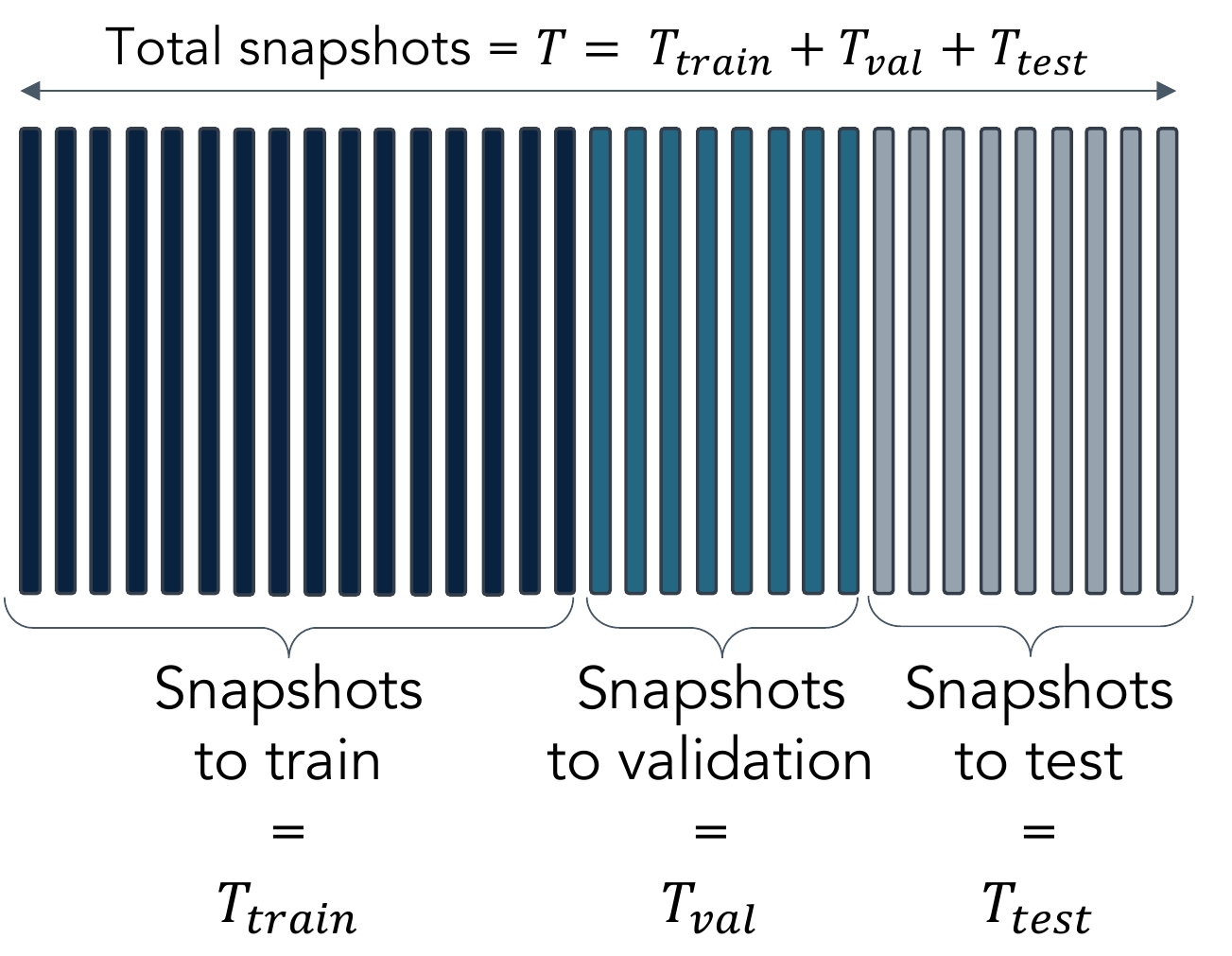}
	\caption{Snapshot organization sketch for time prediction. \label{fig:time_predict}}
\end{figure}

\begin{table}[H]
\caption{Number of snapshots for training, validation and test sets for each one of the databases analysed.}
\centering
\begin{tabular}{ccccc}
\toprule[0.8mm]
\multicolumn{1}{c|}{\textbf{Database}} & \multicolumn{1}{c|}{$T_{train}$} & \multicolumn{1}{c|}{$T_{val}$} & \multicolumn{1}{c|}{$T_{test}$} & \multicolumn{1}{c}{T} \\ 
\midrule \midrule
\multicolumn{1}{c|}{\textbf{ABL}} & \multicolumn{1}{c|}{72} & \multicolumn{1}{c|}{13} & \multicolumn{1}{c|}{15} & \multicolumn{1}{c}{100} \\ 
\hline
\multicolumn{1}{c|}{\textbf{Cyl 2D}} & \multicolumn{1}{c|}{68} & \multicolumn{1}{c|}{12} & \multicolumn{1}{c|}{71} & \multicolumn{1}{c}{151} \\ 
\hline
\multicolumn{1}{c|}{\textbf{Cyl 3D}} & \multicolumn{1}{c|}{170} & \multicolumn{1}{c|}{30} & \multicolumn{1}{c|}{99} & \multicolumn{1}{c}{299} \\ 
\bottomrule[0.8mm]
\end{tabular}
\label{tab:snap}
\end{table}
 
The open source platform known as \textit{TensorFlow} has been used to design the network. Specifically, the \textit{Keras} programming interface included in \textit{TensorFlow} has been used.

\smallskip

\section{Results and Discussions \label{sec:results}}

This section will show in detail the performance of the methodology explained in Sec. \ref{sec:meth} applied to each one of the different databases. 


\subsection{Atmospheric Boundary Layer}
 
 When applying the method with neural networks to the ABL database, the results have been separated into two different tests. On test 1, the downsampling has been performed equispaced, while on test 2, the downsampling has been non-homogeneous, choosing more points on the area near the wall, located at low $y$ values. In addition, the scaling of the data has also been tested.
 Once the SVD is applied, an additional dimension for the time is created but the same snapshot is repeat all the time, to enlarge data dimensionality. 
 Then, the three input tensors are then separated into the training, validation and test sets. Afterwards, the ones selected for the training part are introduced into the network and after an iterative process the best weights that define this problem are obtained. In the test, the reduced flow field is reconstructed. The number of snapshots selected for the training, validation and test datasets are indicated in Tab. \ref{tab:snap}.


 Test 1 corresponds to the results of the reconstruction when the sparse data are obtained homogeneously. Tab.\ref{tab:ABLdownMSE} shows the values of the MSE according to different downsampling. It can be seen that the best results are obtained for a downsampling of 1:50 and that increasing the number of points on the reduced tensor does not improve the error suggesting that overfitting is ocurring in this case. This is because the neural network architecture is developed to work on extreme conditions (very sparse data), representing realistic and industrial problems. However, it can be seen that for a downsampling of 1:90 the same error in pressure and temperature is achieved as for a downsampling of 1:50. So, there is an optimal performance for this model.
 
 \begin{table}[H]
     \caption{MSE error obtained in different homogeneous downsampling for the ABL database.}
        \centering
        \resizebox{14cm}{!} {
        \begin{tabular}{ c | m{2.5cm} | r | r | r | r }
        \toprule[0.8mm]
        \multicolumn{2}{m{3.8cm}|}{\textbf{Downsampling}} & \textbf{Streamwise velocity}& \textbf{Pressure} & \textbf{Temperature} & \textbf{TOTAL} \\
        \midrule \midrule
        1:10 & \centering No scaling &0.012 & 0.052 & 0.000006 &  0.021\\
        \includegraphics[trim= 0mm 0mm 0mm 0mm, clip, width=30mm]{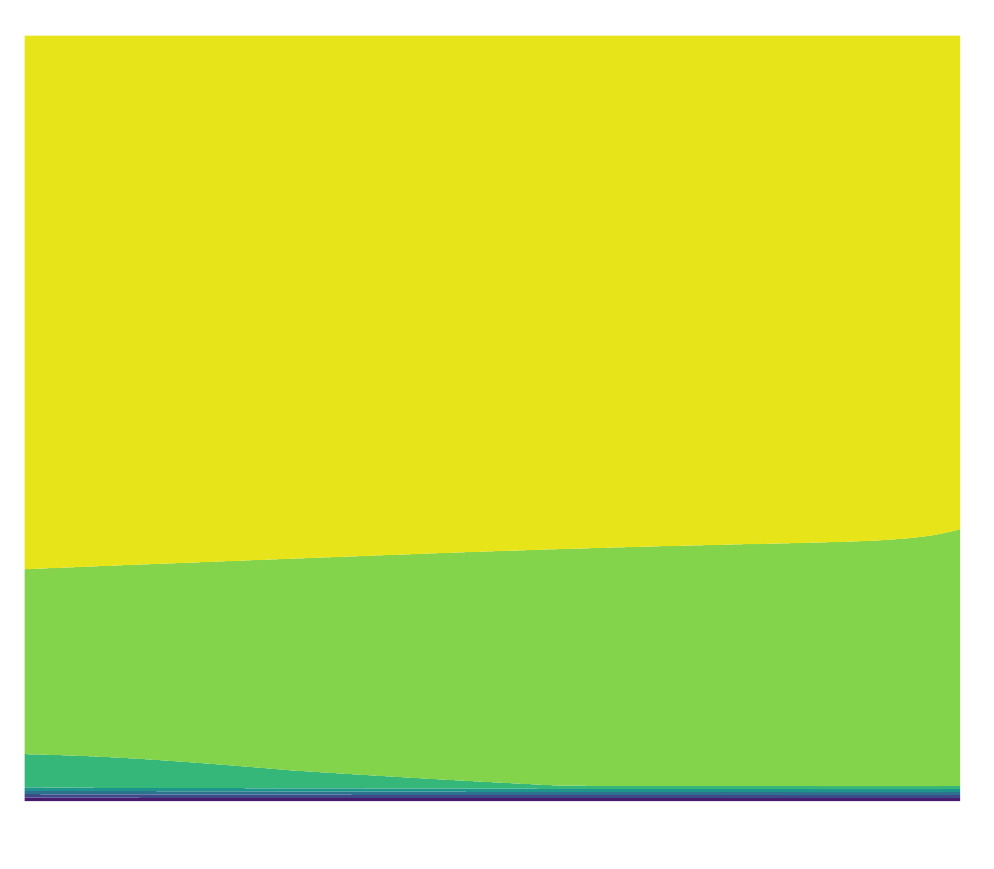} & \centering Scaling & 0.00031 &0.30  &0.000001  & 0.10 \\
        \hline
        1:50 & \centering No scaling & 0.00033 & 0.052 & 0.000006 & 0.017  \\
        \includegraphics[trim= 0mm 0mm 0mm 0mm, clip, width=30mm]{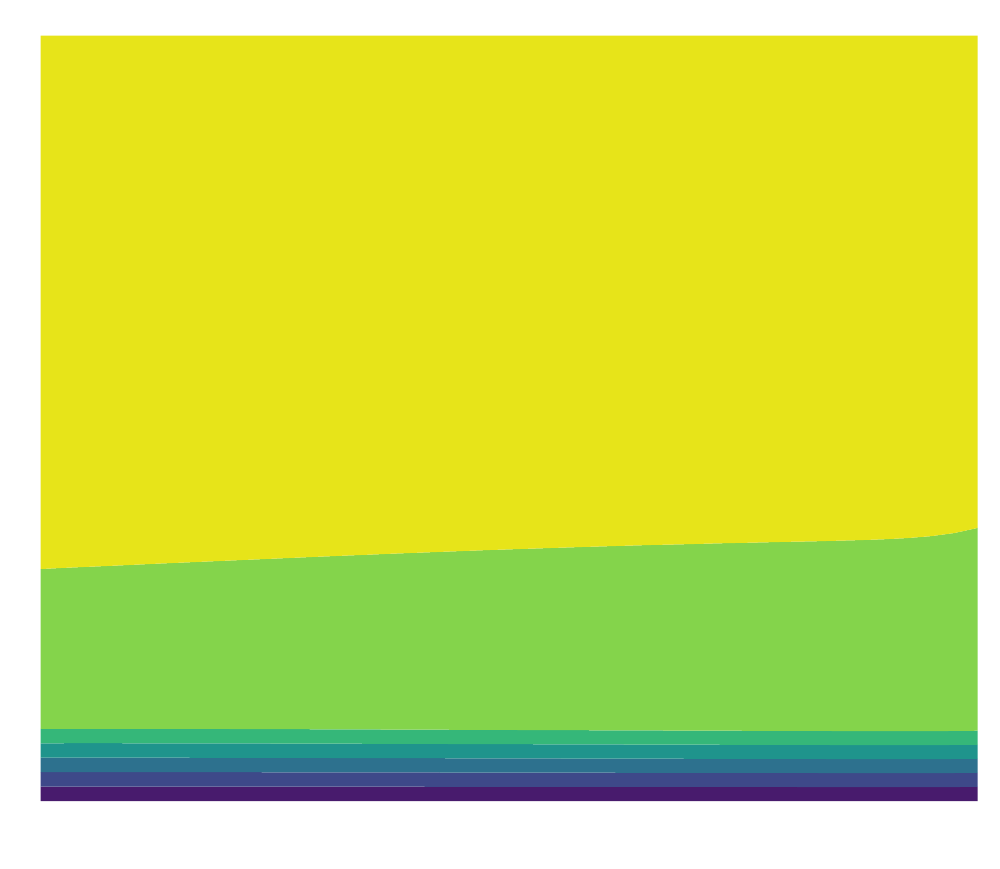} & \centering Scaling & 0.00034 & 0.30 &0.000001  &0.10  \\
        \hline
        1:90 & \centering No scaling &0.0019 &0.052 &0.000006 &0.018 \\
        \includegraphics[trim= 0mm 0mm 0mm 0mm, clip, width=30mm]{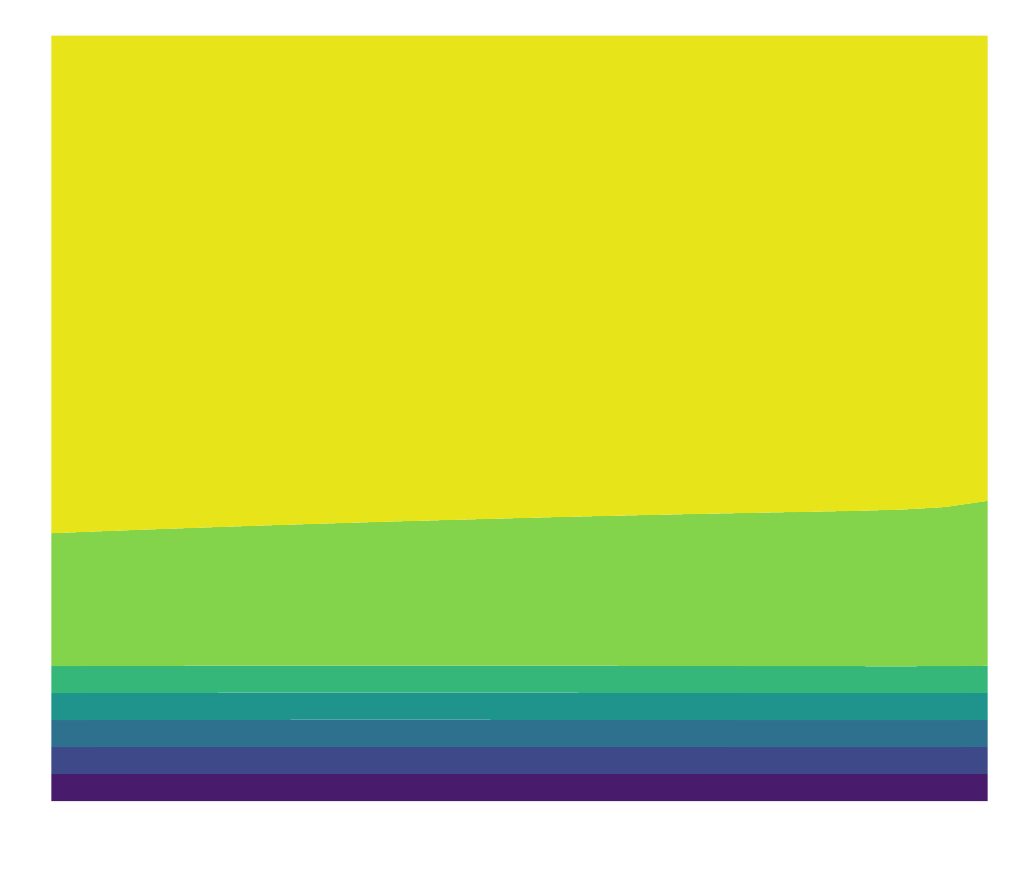} & \centering Scaling &0.012 &0.30 &0.000005 &0.10 \\
       \bottomrule[0.8mm]
       \end{tabular}
       }
     \label{tab:ABLdownMSE}
\end{table}

\begin{figure}[H]
	\centering
	\includegraphics[width = 14 cm]{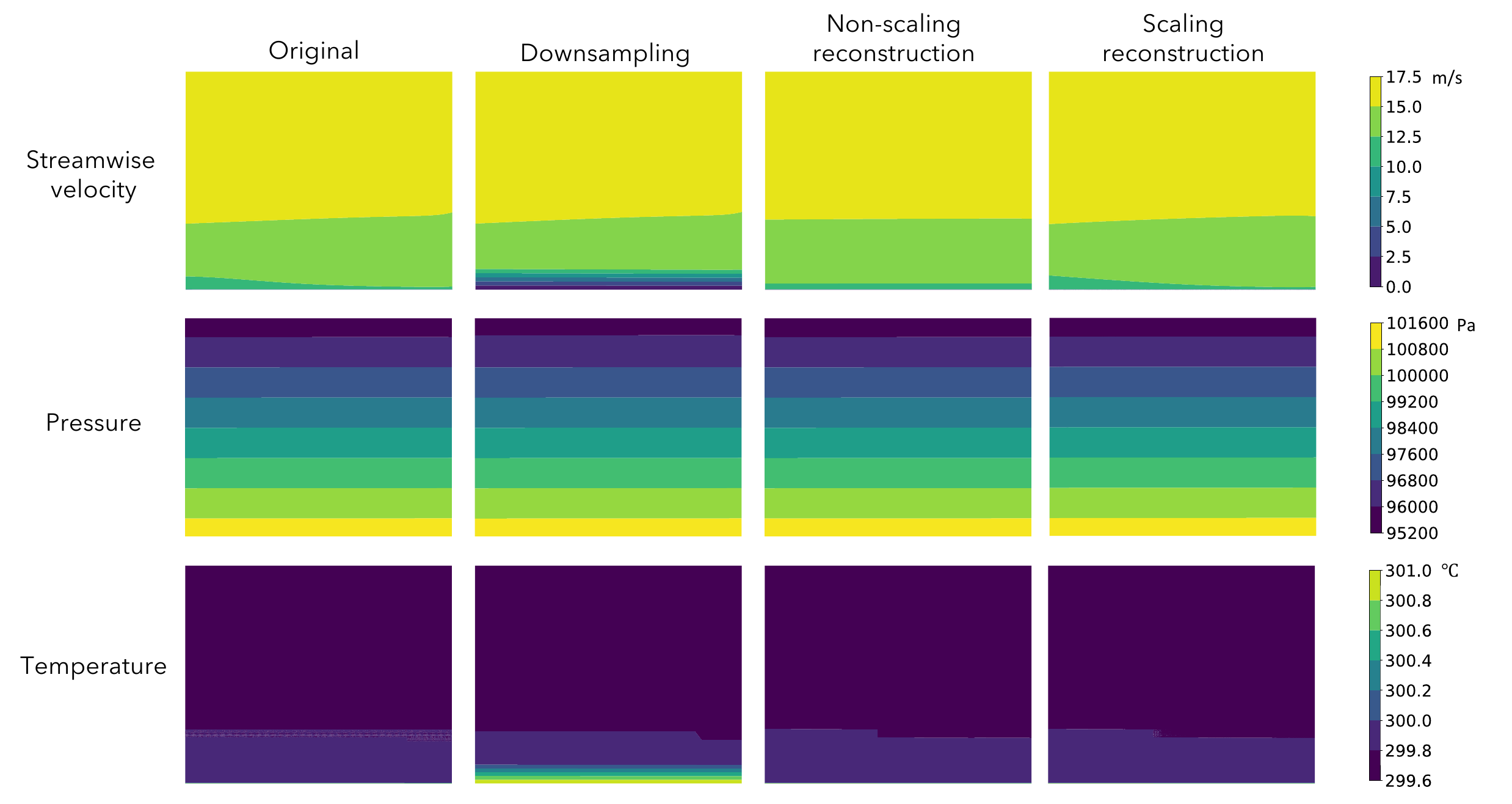}
	\caption{Results of the ABL base neural network method performing 1:50 homogeneous downsampling. The horizontal axis indicates the distance in meters from the study field [0,5000] and the vertical axis the height in meters [0,500].\label{fig:ABLNN11}}
\end{figure}

In Fig. \ref{fig:ABLNN11} the results obtained applying a downsampling of 1:50 are shown. On the third column we present the results without scaling the data and on the fourth column, the results by scaling the data. Comparing the representation of the reconstructed tensor with the original tensor, there are hardly any visual differences. It is also observed that the loss of information that occurs in the velocity variable in the reduced tensor (blue band), is recovered in the reconstructed tensor (the blue band disappears). The same happens with the yellow stripe in the temperature variable.

\begin{figure}[H]
	\centering
	\includegraphics[width = 14 cm]{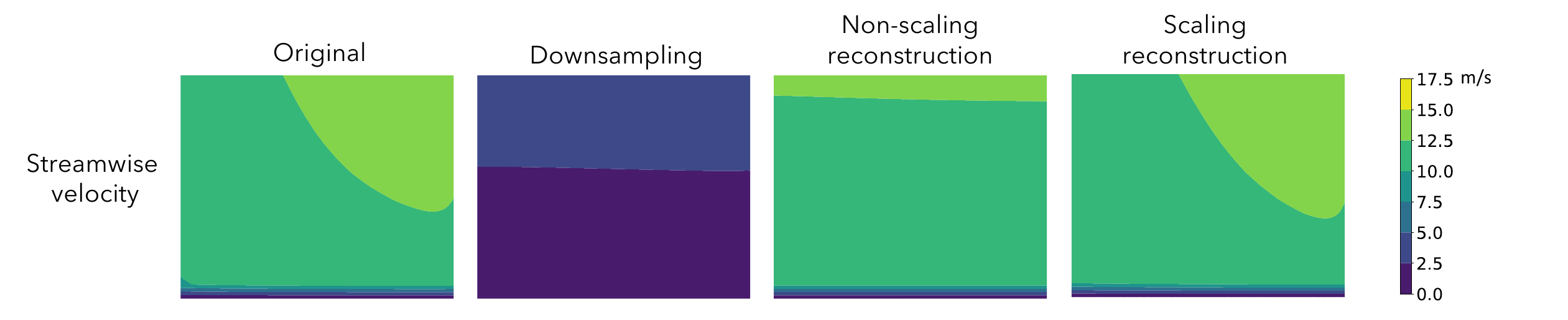}
	\caption{Same as figure \ref{fig:ABLNN11} but the vertical axis dimensions are [0,15] meters, near the wall.\label{fig:ABLNN11zoom}}
\end{figure}

  A zoomed-in view of the area near the wall (from 0 to 15 meters) can be seen in \ref{fig:ABLNN11zoom}. The downsampling representation of the velocity indicates that the velocity is zero in this region. This occurs because the points chosen for downsampling correspond as the zero velocity points in the original representation. Nevertheless, this method is capable to reconstruct this region with high accuracy.

 Test 2 shows the results of performing a non-homogeneous downsampling to the original tensor of the ABL database. More points are taken in the lower zone of the boundary layer, where the streamwise velocity shows a greater contrast.
 
\begin{figure}[H]
	\centering
	\includegraphics[width = 14 cm]{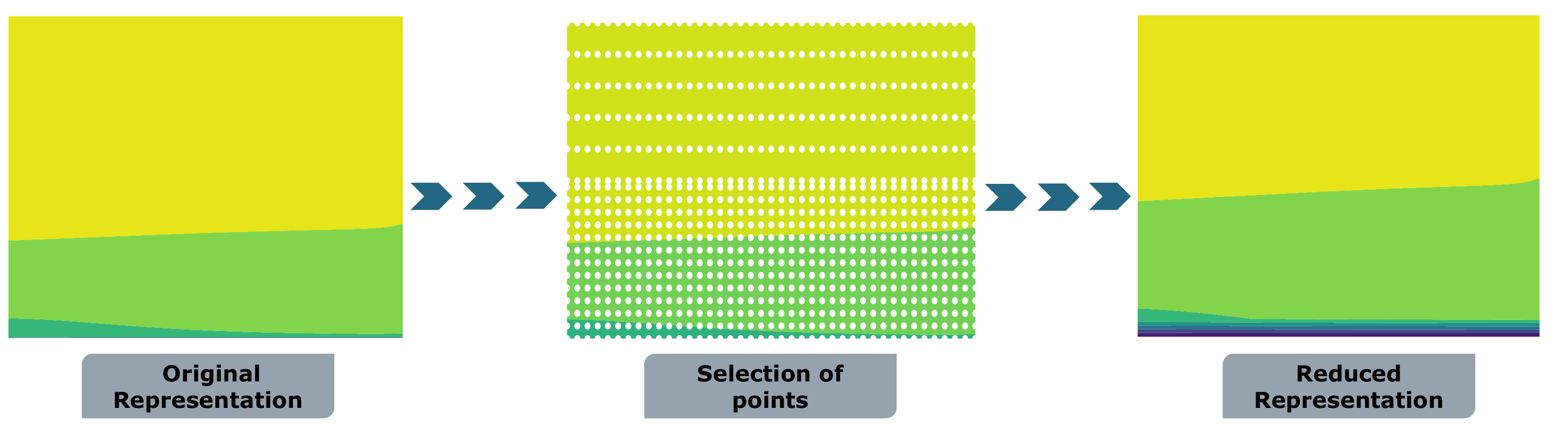}
	\caption{Sketch of the non-homogeneous downsampling performed on the ABL database. \label{fig:esquema_ABL_NN_down}}
\end{figure}

 The results presented correspond to a downsampling where in the $x$ dimension 1 point out of 50 points has been chosen but in the $y$ dimension it has been divided into two equal parts: in the first part from 0 m to 250 m 1 point out of 20 points has been chosen and in the second part from 250 m to 500 m 1 point out of 50 points has been chosen. The result provides a reduced tensor of reduced dimensions $3 \times 19 \times 41$ where there are 41 points in $x$ and 19 points in $y$. In Fig. \ref{fig:esquema_ABL_NN_down}, the selected points are displayed graphically.
 
\begin{figure}[H]
	\centering
	\includegraphics[width = 14 cm]{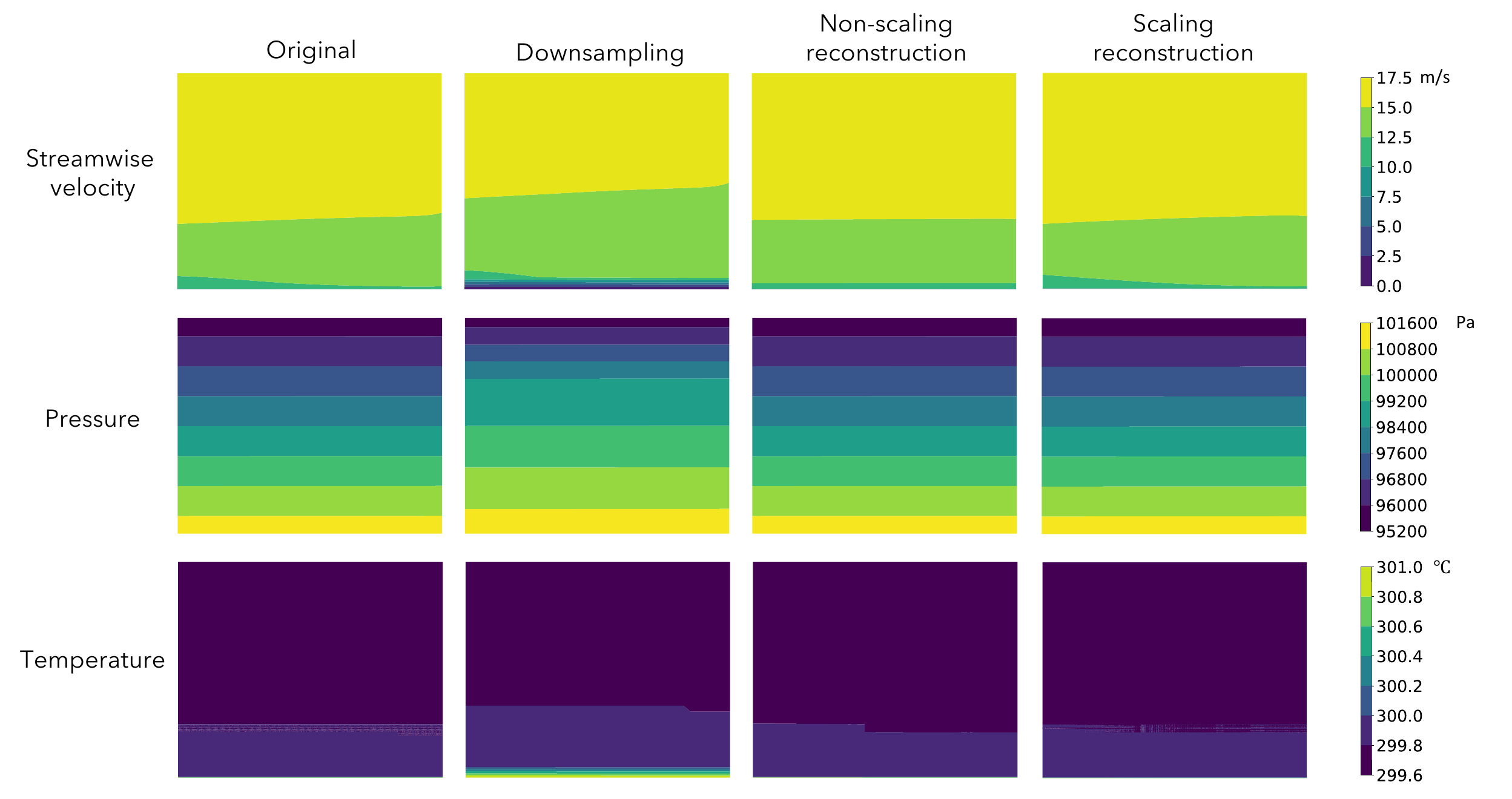}
	\caption{Results of the neural network method on the basis of ABL performing non-homogeneous downsampling. The horizontal axis indicates the distance in meters from the study field [0,5000] and the vertical axis the height in meters [0,500].\label{fig:ABLNN21}}
\end{figure}

The results obtained from the neural network with non-homogeneous downsampling are shown in Fig. \ref{fig:ABLNN21}. The differences with the test 1 are barely noticeable on most of the domain. The only appreciable difference is in the reduced representations of velocity and temperature, second column of Fig. \ref{fig:ABLNN21}. Due to the fact that more points have been chosen in the lower zone, the zero velocity band (blue band) and the yellow band in temperature have been reduced. 

Finally, Tab.\ref{tab:ABLcomparacionRRMSE} and \ref{tab:ABLcomparacionMSE} summarizes and compares the results obtained for the ABL database comparing the RRMSE and MSE in all tests performed, respectively. In Tab.\ref{tab:ABLcomparacionRRMSE} the computational cost is also included. 

\begin{table}[H]
     \caption{Percentage of RRMSE error committed when constructing the ABL database and computational cost ($t_c$). Test 1 refers to homogeneous downsampling results (1:50) and Test 2 refers to non-homogeneous downsampling results (Below: 1:20; Above: 1:50).}
        \centering
        \resizebox{14cm}{!} {
        \begin{tabular}{ c | c | r | r | r | r | r}
        \toprule[0.8mm]
        \multicolumn{2}{m{3.8cm}|}{\textbf{Method}} & \textbf{Streamwise velocity} & \textbf{Pressure} & \textbf{Temperature} & \textbf{TOTAL}  & \textbf{$t_c$(s)} \\
        \midrule \midrule
        \multirow{2}{3cm}{\centering NN Test 1} & No scaling & 0.12 & 0.00023 & 0.00081 & 0.00023& 93.9
        \tabularnewline \cline{2-2} & Scaling & 0.12 & 0.00039 & 0.00056 & 0.00056&80.1 \\
        \hline
        \multirow{2}{3cm}{\centering NN Test 2} & No scaling & 0.73 & 0.00023 & 0.00081 & 0.00026 & 99.3
        \tabularnewline \cline{2-2} & Scaling & 0.11 & 0.00095 & 0.00044 & 0.00095 & 99.1\\
       \bottomrule[0.8mm]
       \end{tabular}
       }
     \label{tab:ABLcomparacionRRMSE}
\end{table}

\begin{table}[H]
     \caption{MSE error made when constructing the ABL database. Test 1 refers to homogeneous downsampling results (1:50) and Test 2 refers to non-homogeneous downsampling results (Below: 1:20; Above: 1:50).}
        \centering
        \resizebox{14cm}{!} {
        \begin{tabular}{ c | c | r | r | r | r }
        \toprule[0.8mm]
        \multicolumn{2}{m{3.8cm}|}{\textbf{Method}} &\textbf{Streamwise velocity} & \textbf{Pressure} & \textbf{Temperature} & \textbf{TOTAL}\\
        \midrule \midrule
        \multirow{2}{3cm}{\centering NN Test 1} & No scaling & 0.00034  & 0.052 & 0.000006 & 0.017
        \tabularnewline \cline{2-2} & Scaling & 0.0003 & 0.30 & 0.000005 & 0.10 \\
        \hline
        \multirow{2}{3cm}{\centering NN Test 2} & No scaling & 0.012 & 0.052 & 0.000006 & 0.021
        \tabularnewline \cline{2-2} & Scaling & 0.0003 & 0.88 & 0.000002 & 0.29  \\
       \bottomrule[0.8mm]
       \end{tabular}
       }
     \label{tab:ABLcomparacionMSE}
\end{table}

Regarding the need of scaling the data, the results show that there is no significant improvement in terms of errors either. However, scaling the data reduces the computational cost because the data are less dispersed. Therefore, the choice of whether or not to scale the data will depend on which objective is more important: computational cost or to achieve a lower error in certain variables. Finally, a comparison of test 1 with test 2 leads to the conclusion that the non-homogeneous downsampling applied improves the results very slightly but at the computational cost increases.


\subsection{Two-dimensional flow past a circular cylinder}

The methodology for this new database is the same as the one applied to the ABL database, although in this case the flow is unsteady. As a reminder, the network architecture is exactly the same as indicated in Sec. \ref{sec:meth} showing the robustness and the generalization of the model presented.

For this database, different results are presented varying the number of points chosen for downsampling. Also, tests have been performed scaling the input data. A summary of the MSE obtained in the different tests is shown in Tab. \ref{tab:Cil2DdownMSE}. The table shows the representation of a representative snapshot of the reduced matrix for the U velocity component. The original tensor of this database has dimensions of $3 \times 449 \times 199 \times 151$ so performing a 1:50 downsampling gives a reduced tensor of $3 \times 9 \times 4 \times 151$ and a downsampling of 1:10 one tensor of $3 \times 45 \times 20 \times 151$. 

The table shows that for the case without scaling, the total error decreases as more points are introduced into the network. However, this trend is not observed when the data is scaled since for a downsampling of 1:10 the error shoot up. This suggest that \textit{overfitting} is ocurring: too much data is introduced into the network. This is related to both the number of data and the dispersion of those data, so it occurs when the data are scaled and not when they are not scaled.

\begin{table}[H]
     \caption{MSE error obtained in different downsampling for the two-dimensional cylinder database.}
        \centering
        \resizebox{14cm}{!} {
        \begin{tabular}{ c | m{2.5cm} | r | r | r | r }
        \toprule[0.8mm]
        \multicolumn{2}{m{3.8cm}|}{\textbf{Downsampling}} & \textbf{Velocity U}& \textbf{Velocity V} & \textbf{Vorticity} & \textbf{TOTAL} \\
        \midrule \midrule
        1:10 & \centering No scaling & 0.011 & 0.0023 & 0.033 & 0.016 \\
        \includegraphics[trim= 0mm 0mm 0mm 0mm, clip, width=30mm]{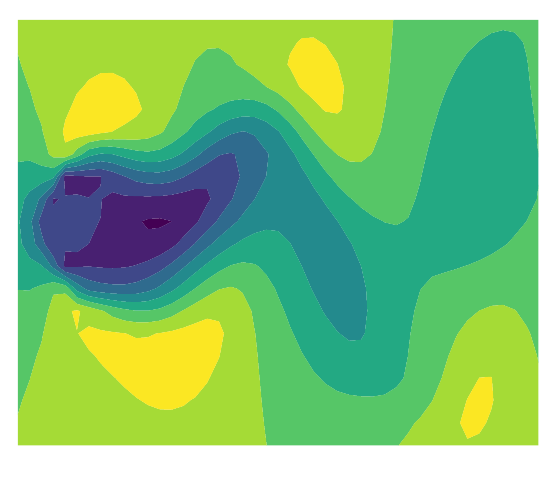} & \centering Scaling & 0.019 & 0.010 & 4.26 & 1.43 \\
        \hline
        1:20 & \centering No scaling & 0.011 & 0.0024 & 0.063 & 0.025 \\
        \includegraphics[trim= 0mm 0mm 0mm 0mm, clip, width=30mm]{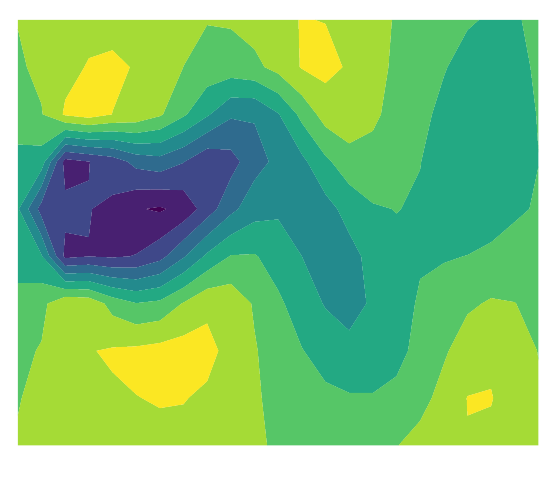} & \centering Scaling & 0.0021 & 0.0017 & 0.40 & 0.14 \\
        \hline
        1:30 & \centering No scaling & 0.0037 & 0.00073 & 0.10 & 0.035 \\
        \includegraphics[trim= 0mm 0mm 0mm 0mm, clip, width=30mm]{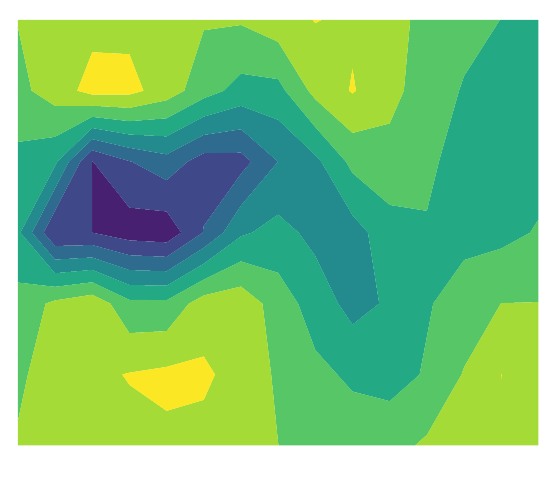} & \centering Scaling & 0.00086 &0.00069  &0.16  & 0.054 \\
        \hline
        1:40 & \centering No scaling &0.0059 &0.0019 &0.19 &0.068 \\
        \includegraphics[trim= 0mm 0mm 0mm 0mm, clip, width=30mm]{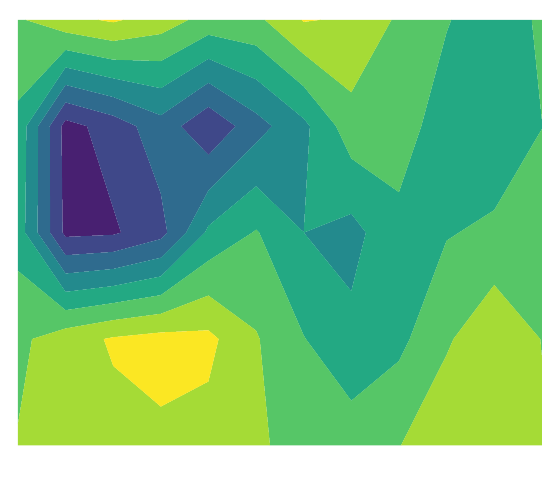} & \centering Scaling &0.0022  &0.0015  &0.22  & 0.076 \\  
        \hline
        1:50 & \centering No scaling & 0.0016&0.0057 &0.31 & 0.11\\
        \includegraphics[trim= 0mm 0mm 0mm 0mm, clip, width=30mm]{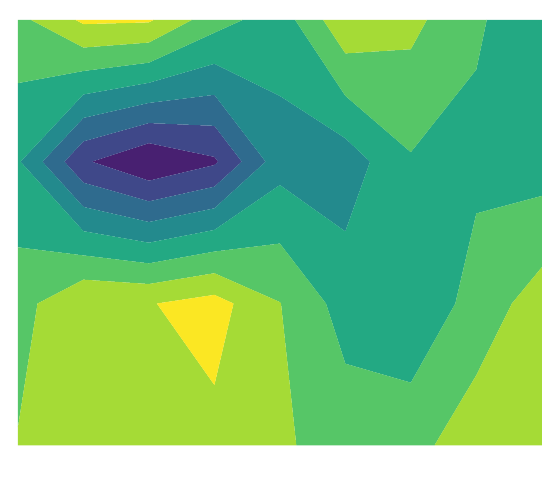} & \centering Scaling & 0.0050 & 0.0034 & 0.36 & 0.12 \\
       \bottomrule[0.8mm]
       \end{tabular}
       }
     \label{tab:Cil2DdownMSE}
\end{table}

To show that the error for the 1:10 case with scaling is triggered due to overfitting, the number of modes of the SVD decomposition that will enter the network has been reduced. The comparison of the MSE obtained for a downsampling of 1:10 with and without a mode reduction is shown in Tab. \ref{tab:Cil2DdownMSElimm}. As seen, reducing the number of modes, i.e., reducing the amount of data introduced on the neural network, improves the performance of the interpolation.

\begin{table}[H]
     \caption{MSE error committed for a downsampling of 1:10 by reducing the number of modes to 10 in the tow-dimensional cylinder database.}
        \centering
        \resizebox{14cm}{!}{
        \begin{tabular}{ c | m{2.5cm} | r | r | r | r }
        \toprule[0.8mm]
        \multicolumn{2}{m{3.8cm}|}{\textbf{Downsampling}} & \textbf{Velocity  U}& \textbf{Velocity  V} & \textbf{Vorticity} & \textbf{TOTAL} \\
        \midrule \midrule
        1:10 & \centering Mode reduction & 0.0035 & 0.0042 & 0.71 & 0.24 \\
        \includegraphics[trim= 0mm 0mm 0mm 0mm, clip, width=25mm]{cyl2d_u_down_10.pdf} & \centering No mode reduction & 0.019 & 0.010 & 4.26 & 1.43 \\
             \bottomrule[0.8mm]
       \end{tabular}
       }
     \label{tab:Cil2DdownMSElimm}
\end{table}

\begin{figure}[H]
	\centering
	\includegraphics[width = 14 cm]{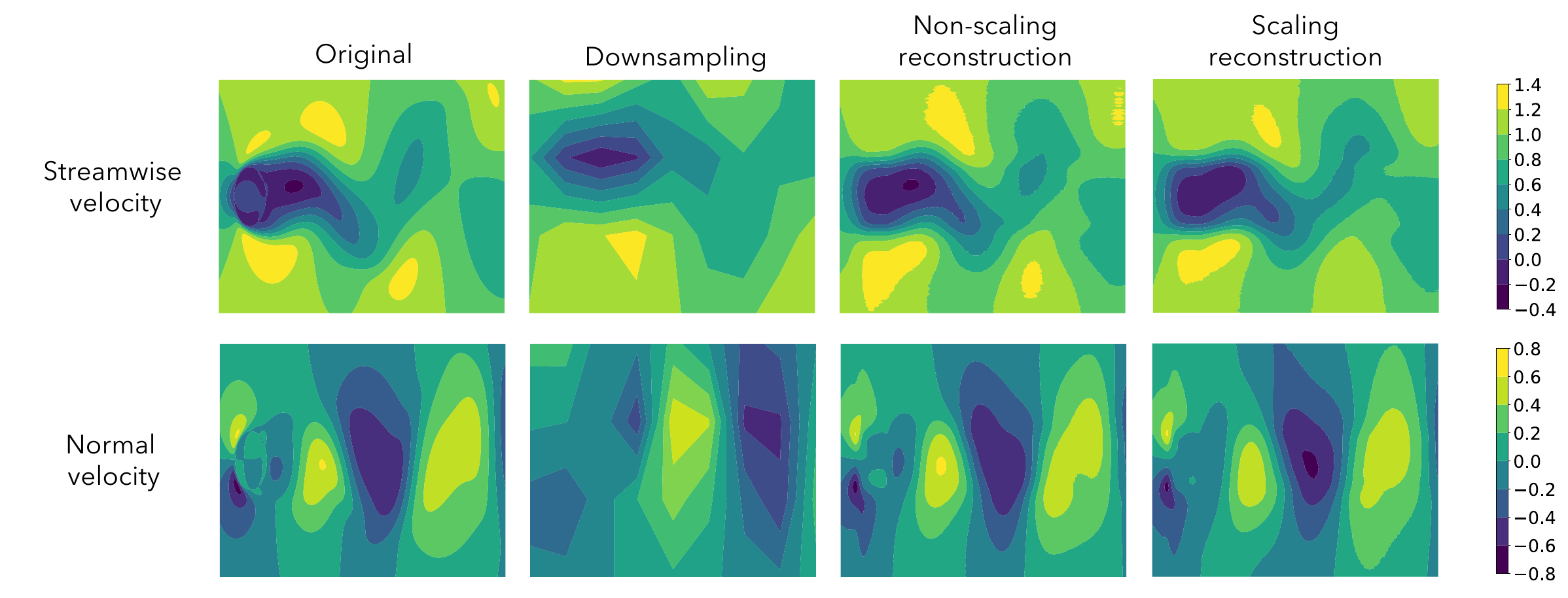}
	\caption{Results of the neural network method of two-dimensional cylinder base data by performing 1:50 downsampling. The horizontal coordinate axis indicates the horizontal distance between [-1,8] and the vertical axis indicates the vertical distance between [-2,2]. Snapshot number = 80. \label{fig:Cil2DNN11}}
\end{figure}

 Finally, graphical representations of the results for the most extreme downsampling (1:50) are shown in Fig. \ref{fig:Cil2DNN11}. The input to the network are represented on the second column of Fig. \ref{fig:Cil2DNN11}, while the reconstructions are represented on the third column for non scaling data and on the fouth column for scaling data. As seen, some differences can be appreciated between the reconstructions and the original snapshots (first column). The reconstruction of points has been carried out on the three variables that make up this database ($v_x$, $v_y$ and vorticity), however, it has been decided to analyze the two most important physical variables ($v_x$ and $v_y$) because with the composition of these variables the vorticity can be obtained.

The results are then compared using error analysis. Tab. \ref{tab:Cil2DcomparacionMSE} shows the results of the MSE for a downsampling of 1:50. Due to use the MSE as loss function, an attempt is made to optimize this error. If a comparison is made between the results with scaling and without it, the scaling results provides better results.
 
The results corresponding to the RRMSE are shown in Tab. \ref{tab:Cil2DcomparacionRRMSE}, as well as an average of the computational cost of each algorithm. These errors are higher than the MSE seen above because they are triggered by making errors when reconstructing data whose value is 0, although it is still interesting to analyze as it provides the percentage of data that could not be reconstructed. In the same way than before, the scaling results are around 5\% order better than no scaling ones.

Finally, 5\% and 10\% of random noise has been added to the original database to test the response of the NN architecture in more realistic situations (as in the case of experimental databases). These results show that the error is higher than the tests without noise but remains within  similar values as in the previous cases, but increasing the error with a level similar to the noise range.

In terms of computational cost, the cost of interpolations is lower than that of NNs, but once the model, which is robust, is developed, the computational cost of NNs is still low (less than 1 min) and provides better results than the simple interpolation. 

\begin{table}[H]
     \caption{MSE error committed when reconstructing the two-dimensional cylinder database with 1:50 downsampling.}
        \centering
        \resizebox{12cm}{!} {
        \begin{tabular}{ c | c | r | r | r }
        \toprule[0.8mm]
        \multicolumn{2}{m{3.8cm}|}{\textbf{Method}} & \textbf{Velocity  U}& \textbf{Velocity  V} & \textbf{TOTAL} \\
        \midrule \midrule
        \multirow{2}{3cm}{\centering NN} & No scaling & 0.016 & 0.0057 & 0.011
        \tabularnewline \cline{2-2} & Scaling & 0.0050 & 0.0034 & 0.0042 \\
        \hline
        \multirow{2}{3cm}{\centering NN noise test} & 5\% interference & 0.025 & 0.0015 & 0.013
        \tabularnewline \cline{2-2} & 10\% interference & 0.030 & 0.018 & 0.024 \\
       \bottomrule[0.8mm]
       \end{tabular}
       }
     \label{tab:Cil2DcomparacionMSE}
\end{table}

\begin{table}[H]
     \caption{Percentage of RRMSE error committed when reconstructing the two-dimensional cylinder database with 1:50 downsampling and computational cost.}
        \centering
        \resizebox{12cm}{!} {
        \begin{tabular}{ c | c | r | r | r | r}
        \toprule[0.8mm]
        \multicolumn{2}{m{3.8cm}|}{\textbf{Method}} & \textbf{Velocity  U}& \textbf{Velocity  V} & \textbf{TOTAL} & \textbf{$t_c$ (s)} \\
        \midrule \midrule
        \multirow{2}{3cm}{\centering NN} & No scaling & 14.1 & 29.0 & 21.5 & 54.4
        \tabularnewline \cline{2-2} & Scaling & 7.9& 22.4 & 15.1 & 55.4 \\
        \hline
        \multirow{2}{3cm}{\centering NN noise test} & 5\% interference & 17.2 & 46.9 & 32.5 & 20.4
        \tabularnewline \cline{2-2} & 10\% interference & 18.1& 51.0 & 34.5 & 22.6 \\
       \bottomrule[0.8mm]
       \end{tabular}
       }
     \label{tab:Cil2DcomparacionRRMSE}
\end{table}


\subsection{Three-dimensional flow past a circular cylinder}

The last database to be tested is the three-dimensional flow past a circular cylinder. 

Two main tests are going to be presented. The difference between these two test is the change in the loss function. On test 1, the MSE (eq. (\ref{eqn:MSE})) of the validation set is the loss function, while on test 2, the RRMSE (eq. (\ref{eqn:RRMSE})) is going to be tried. 

The network methodology followed for this database is practically identical to the methodology followed with the rest of the databases. The main difference is that the database has $5$ dimensions (variables, time and three space dimensions), while the other cases have just $4$. Therefore, if the same network architecture is to be used, it has been necessary to add an extra step before applying the SVD decomposition. This extra step consists of a reshaping of the $5$-dimensional tensor to a $4$-dimensional tensor. The chosen option has been to join the dimension corresponding to the points in $x$ with the points in $z$: from an original tensor of dimension $3 \times 100 \times 40 \times 64 \times 299$ an extreme downsampling of 1:20 has been performed on the $x$ and $y$ dimensions, obtaining a reduced tensor of dimension $3 \times 5 \times 2 \times 64 \times 299$ and subsequently applying the redimensionalization, it remains a tensor of $3 \times 5 \times 128  \times 299$. To this last tensor, the SVD decomposition is applied and the resulting matrices are those that are introduced in the network.  The output of the network will be a tensor of $3 \times 100 \times 2560 \times 299$ which will have to be resized again to separate the $x$ and $z$ dimensions.

As done previously, a study has been carried out with different downsampling in order to know which amount of data provides the best results. Tab. \ref{tab:Cil3DdownMSE} shows the MSE obtained for each downsampling. The maximum downsampling done in the database has been 1:20, as the original database has only $40$ and $100$ points in the streamwise and normal directions, respectively. With this downsampling the points are reduced to 2 and 5. As seen in Tab. \ref{tab:Cil3DdownMSE}, the total errors are of the same order of magnitude except for those belonging to the 1:10 downsampling with scaling, where the error increases one order of magnitude due to overfitting. The best results are observed for 1:15 downsampling without data scaling, however, the results obtained for 1:20 downsampling are very good considering the amount of data used. Scaling the data does not seem to improve the performance of the neural network.

\begin{table}[H]
     \caption{MSE error obtained in different downsampling for the three-dimensional cylinder database.}
        \centering
        \resizebox{14cm}{!} {
        \begin{tabular}{ c | c | c | r | r | r | r }
        \toprule[0.8mm]
        \multicolumn{3}{m{3.8cm}|}{ \textbf{Downsampling}} & \textbf{Velocity  U}& \textbf{Velocity  V} & \textbf{Velocity W} & \textbf{TOTAL} \\
        \midrule \midrule
        1:10 & \multirow{2}{2cm}{\centering Test 1} & No scaling & 0.012 &0.0090 &0.00043 &0.0071  \\
         & &Scaling &0.028 &0.025 &0.00094 &0.018 \\
        \includegraphics[trim= 0mm 0mm 0mm 0mm, clip, width=30mm]{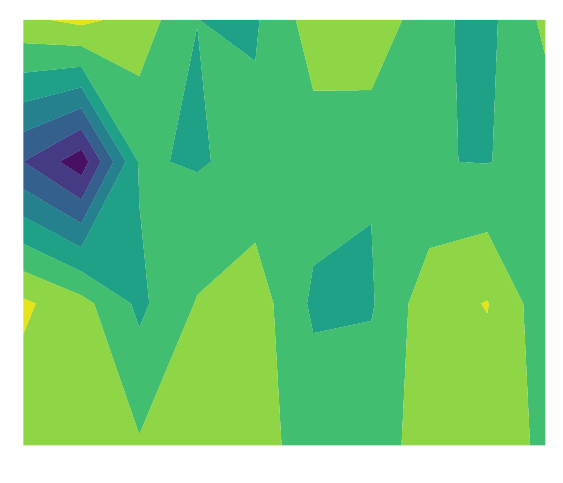} & \multirow{2}{2cm}{\centering Test 2} &No scaling& 0.011 &0.0099  &0.00031 &0.0070  \\
        & &Scaling & 0.022&0.019 &0.00064 &0.014 \\
        \hline
        1:15 & \multirow{2}{2cm}{\centering Test 1} & No scaling &0.0064  & 0.0046& 0.00042& 0.0038 \\
         & &Scaling &0.0092 & 0.0090&0.00054 &0.0062 \\
        \includegraphics[trim= 0mm 0mm 0mm 0mm, clip, width=30mm]{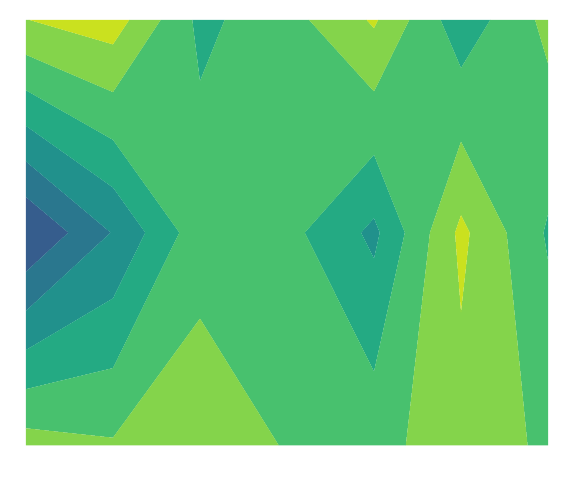} & \multirow{2}{2cm}{\centering Test 2} &No scaling&  0.0051 & 0.0039&0.00035 &0.0031 \\
        & &Scaling &0.0079 &0.0065 &0.00044 &0.0049 \\
        \hline
        1:20 & \multirow{2}{2cm}{\centering Test 1} & No scaling & 0.011 &0.0091 &0.00050 & 0.0070 \\
         & &Scaling &0.0091 &0.0087 &0.00052 &0.0061 \\
        \includegraphics[trim= 0mm 0mm 0mm 0mm, clip, width=30mm]{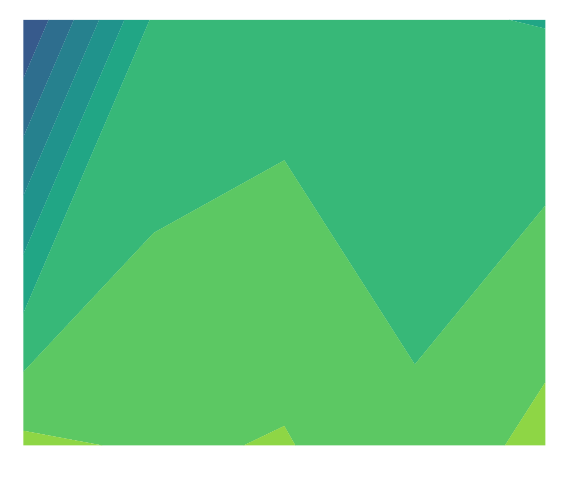} & \multirow{2}{2cm}{\centering Test 2} &No scaling & 0.0095 & 0.0079 &0.00046 &0.0059  \\
        & &Scaling &0.0082 &0.010 &0.00047 &0.0062 \\
       \bottomrule[0.8mm]
       \end{tabular}
       }
     \label{tab:Cil3DdownMSE}
\end{table}

\begin{figure}[H]
	\centering
	\includegraphics[width = 14 cm]{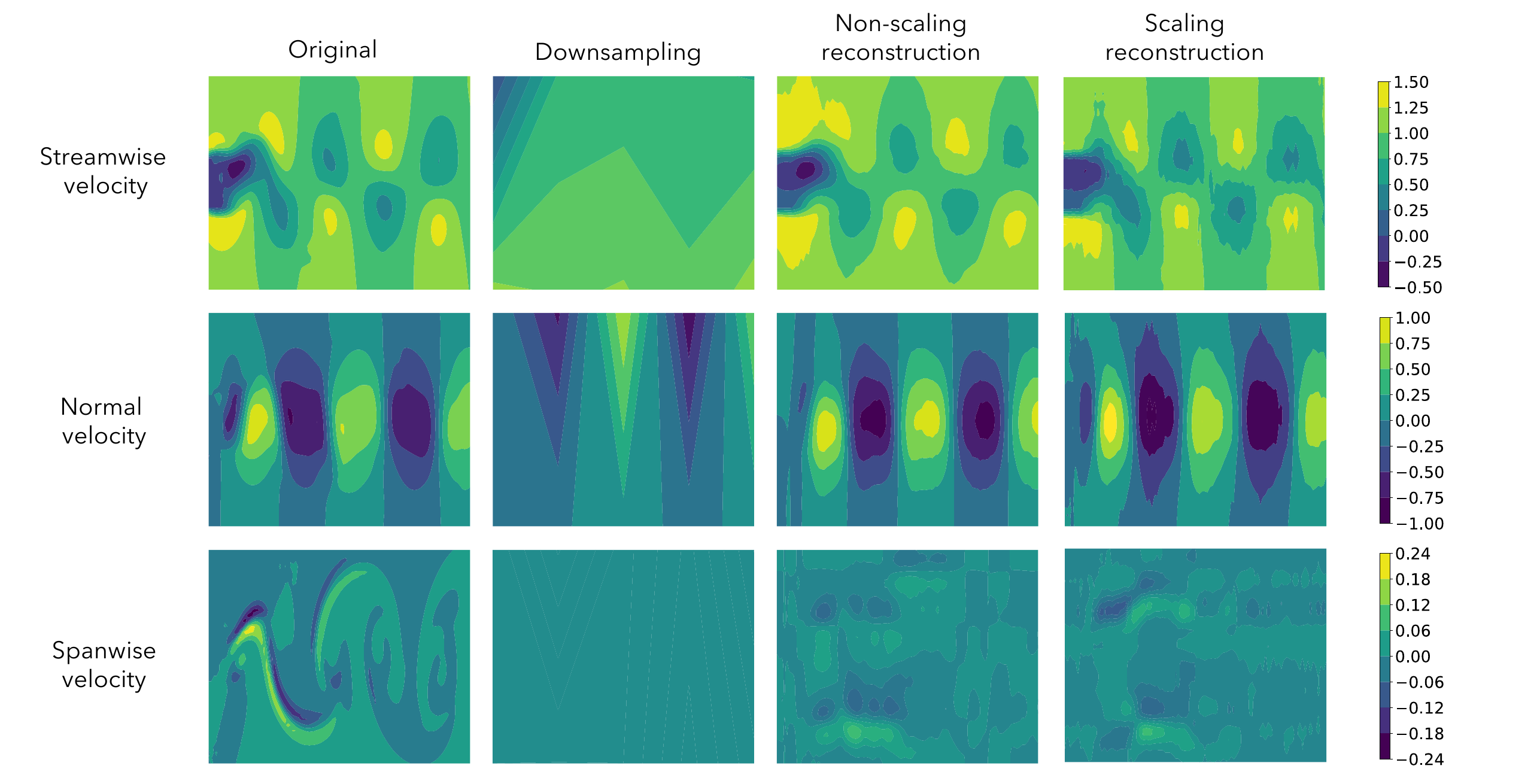}
	\caption{Results of the neural network method of three-dimensional cylinder base data by performing 1:20 downsampling. The horizontal coordinate axis indicates the horizontal distance between [0,10] and the vertical axis indicates the vertical distance between [-2,2]. Snapshot number = 200. z=0.\label{fig:Cil3DNN11}}
\end{figure}

\begin{figure}[H]
	\centering
	\includegraphics[width = 14 cm]{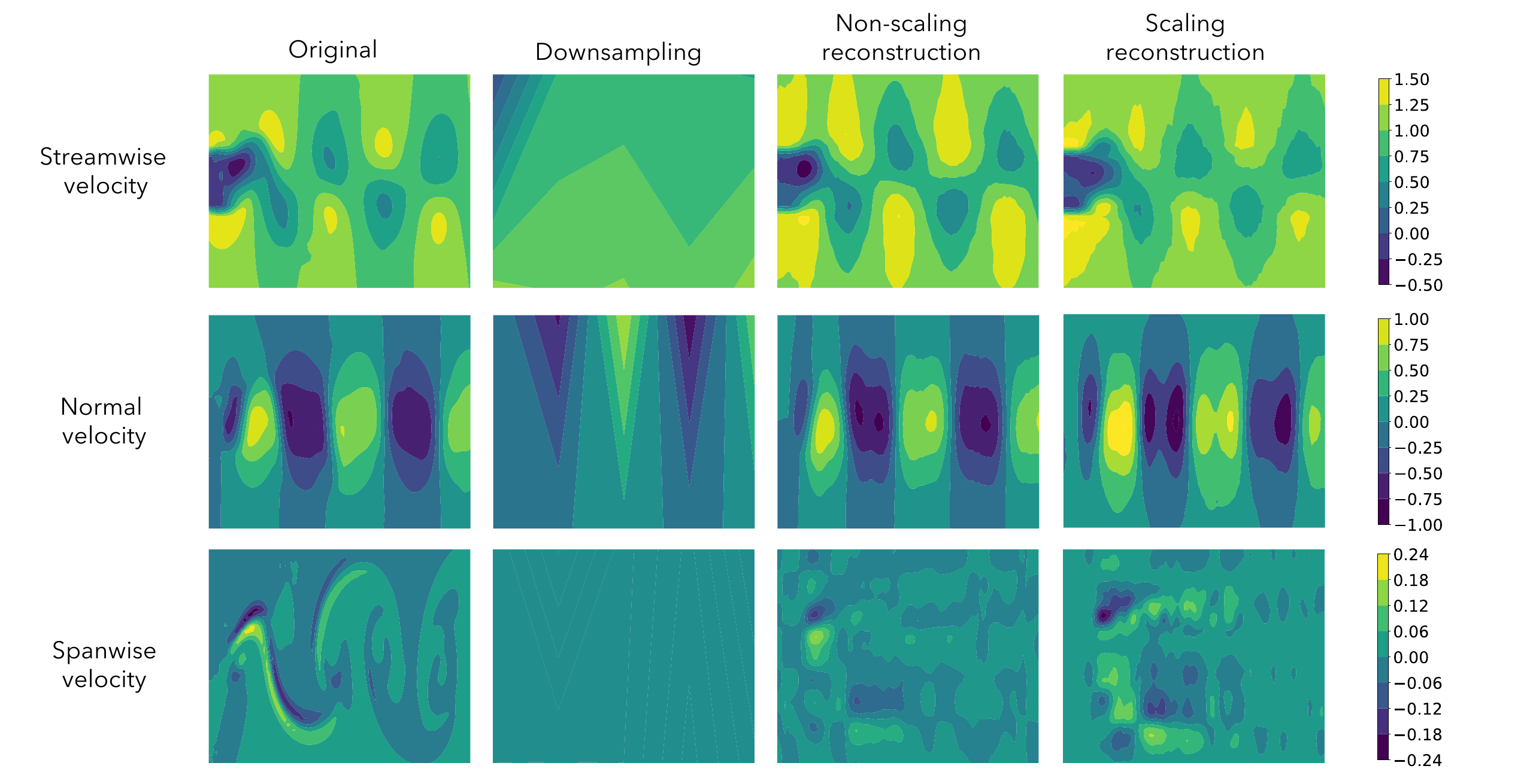}
	\caption{Same as for figure \ref{fig:Cil3DNN11} but for test 2.
\label{fig:Cil3DNN21}}
\end{figure}

Some graphical results from the reconstruction of this database are shown in Figs. \ref{fig:Cil3DNN11} and \ref{fig:Cil3DNN21}. Figure \ref{fig:Cil3DNN11} shows the results for a representative snapshot of test 1. The reduced representations (second column) bear little resemblance to the original representations. However, the reconstructed representations (third column) are really close to the original ones providing very good results considering that 5 points in $x$ and 2 points in $y$ have been introduced. However, In the case of the spanwise velocity (third row), the reconstruction graph is more different than the others compared with the original database. This is due to the fact that most of the $v_z$ velocity terms are zero-valued as can be seen in the original representation. Then, taking into account the above, the reconstruction of these points is associated with a very high relative error. 

The results corresponding to test 2 are shown in Fig. \ref{fig:Cil3DNN21} for the same snapshot and the same z-plane as the graphs presented for test 1. Comparing the results of both tests, no relevant differences can be seen. In the following section, a comparison of errors between the two tests will be made in order to know the differences. Regarding the scaled results that appear in the fourth column of both figures, it is observed that the maximum $U$ velocity points that appear in the original representation (yellow areas) are reconstructed but the values reached are not the correct ones and therefore, the MSE observed in the downsampling table is higher in the cases with scaling.

\begin{figure}[H]
	\centering
	\includegraphics[width = 14 cm]{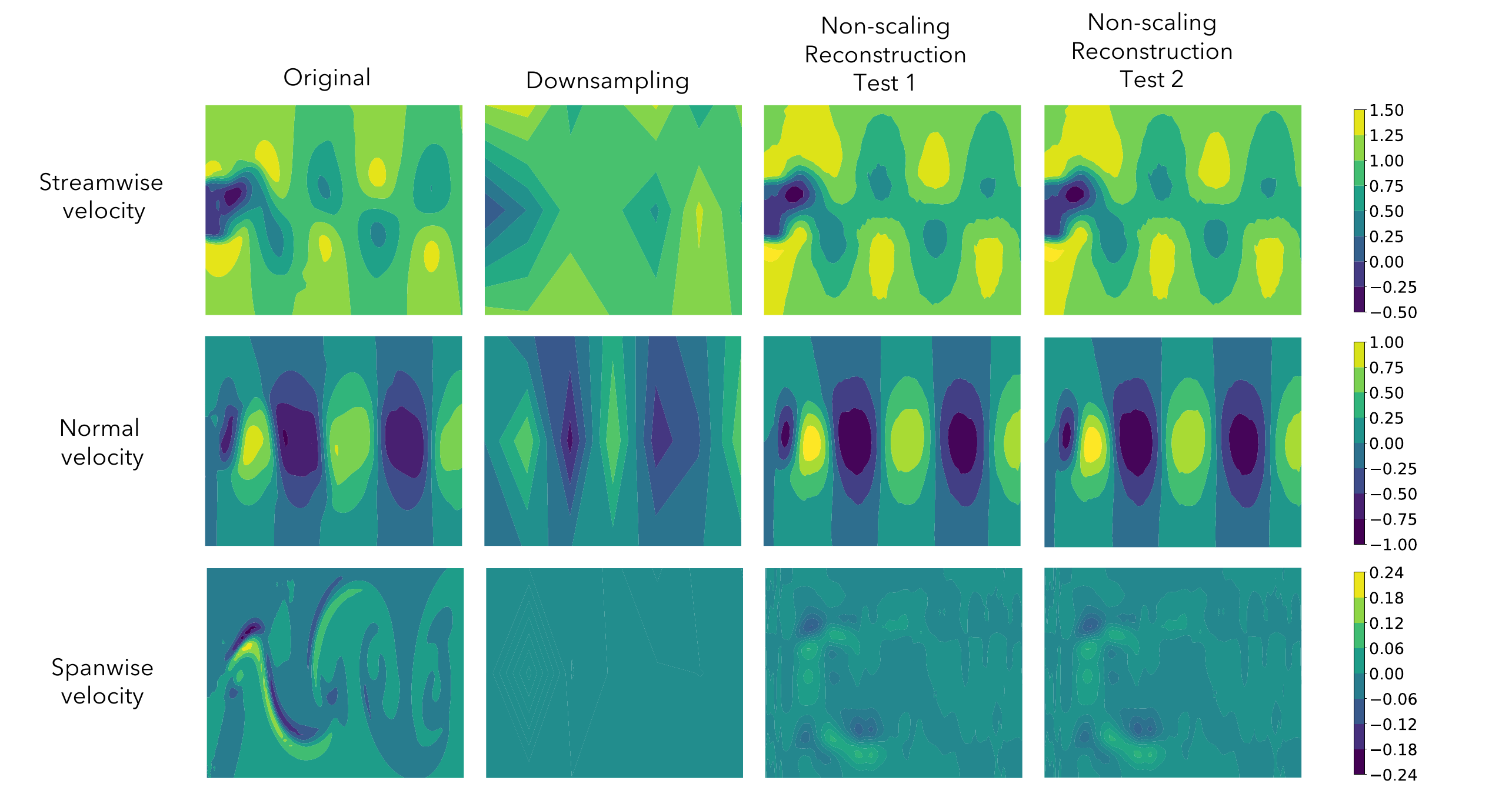}
	\caption{Results of the neural network method of 3D Cylinder base data by performing 1:15 downsampling. The horizontal coordinate axis indicates the horizontal distance between [0,10] and the vertical axis indicates the vertical distance between [-2,2]. Snapshot number = 200. z=0.\label{fig:Cil3DNN}}
\end{figure}

 Fig. \ref{fig:Cil3DNN} includes the results for the optimal downsampling according to the error table discussed above. Increasing the points in $x$ and $y$ from 2 and 5 (downsampling 1:20) to 3 and 7 (downsampling 1:15), the performance of the neural network is highly improved, particularly for the streamwise velocity $U$, where negligible differences can be seen compared to the original snapshot.

Tab. \ref{tab:Cil3DcomparacionMSE} and Tab. \ref{tab:Cil3DcomparacionRRMSE}) shows the MSE and RRMSE reconstruction errors, respectively, for a downsampling of 1:20.

From the MSE Tab. \ref{tab:Cil3DcomparacionMSE} it is observed that the lowest absolute error is obtained for the case of test 2 (loss function RRMSE) without scaling the data. However, the lowest RRMSE is given for the case of test 1 with the data scaled, according to Tab. \ref{tab:Cil3DcomparacionRRMSE}. In addition, it is observed that the RRMSE for the different cases are quite close.
In the case of the spanwise velocity, the RRMSE errors are higher than in the rest of the variables due to the fact that most of the $v_z$ velocity terms are zero and that makes the reconstructed relative error rises. Nevertheless, the MSE error is not as high as the RRMSE error.

The computational cost is higher than for the 2D Cylinder database, as it is more complex. Nevertheless, it is good as the values hardly exceed 5 minutes.

\begin{table}[H]
     \caption{MSE error committed when reconstructing the 3D Cylinder database with a downsampling of 1:20.}
        \centering
        \resizebox{14cm}{!} {
        \begin{tabular}{ c | c | r | r | r | r }
        \toprule[0.8mm]
        \multicolumn{2}{m{3.8cm}|}{\textbf{Method}} &  \textbf{Velocity U} & \textbf{Velocity V} & \textbf{Velocity W} & \textbf{TOTAL} \\
        \midrule \midrule
        \multirow{2}{3cm}{\centering Neural Network Test 1} & Without scaling & 0.011 & 0.0091 & 0.00050 & 0.0070
        \tabularnewline \cline{2-2} & With scaling & 0.0091 & 0.0087 & 0.00052 & 0.0061 \\
        \hline
        \multirow{2}{3cm}{\centering Neural Network Test 2} & Without scaling & 0.0095 & 0.0079 & 0.00046 & 0.0059
        \tabularnewline \cline{2-2} & With scaling & 0.0082 & 0.010 & 0.00047 & 0.0063  \\
       \bottomrule[0.8mm]
       \end{tabular}
       }
     \label{tab:Cil3DcomparacionMSE}
\end{table}

\begin{table}[H]
     \caption{Percentage of RRMSE error committed when reconstructing the 3D Cylinder database with 1:20 downsampling and computational cost.}
        \centering
        \resizebox{14cm}{!} {
        \begin{tabular}{ c | c | r | r | r | r | r}
        \toprule[0.8mm]
        \multicolumn{2}{m{3.8cm}|}{\textbf{Method}} &  \textbf{Velocity U} & \textbf{Velocity V} & \textbf{Velocity W} & \textbf{TOTAL} & \textbf{Cost (s)} \\
        \midrule \midrule
        \multirow{2}{3cm}{\centering Neural Network Test 1} & Without scaling &  11.3 & 28.5 & 89.3 & 14.4 & 169.5
        \tabularnewline \cline{2-2} & With scaling & 10.1 & 27.9  & 90.7 & 12.3 & 300.8 \\
        \hline
        \multirow{2}{3cm}{\centering Neural Network Test 2} & Without scaling & 10.3 & 26.4 & 85.6 & 13.3 &123.8
        \tabularnewline \cline{2-2} & With scaling & 9.5 & 29.9 & 86.3 & 13.6 & 196.6 \\
       \bottomrule[0.8mm]
       \end{tabular}
       }
     \label{tab:Cil3DcomparacionRRMSE}
\end{table}

\section{Conclusions \label{sec:conclusions}}

The present work explores the posibility of using hybrid models, combining physical principles with strategies of machine learning, to reconstruct fluid dynamics databases from sensor measurements. For this purpose, the data dimensionality is reduced using a singular value decomposition (SVD), which extract the main physical patterns modeling the flow. From the reduced database, a very simple and efficient neural network based on autoencoders has been designed to perform to reconstruct a database from a very reduced number of points, which represent sensor measurements. The methodology, which is completely new and simple, has been tested on different databases to check also the robustness of the model developed. More specifically, the method is applied in three different databases representing a turbulent atmospheric boundary layer (ABL) database and the two- and three- dimensional flow past a circular cylinder.
 
The performance of the method has been tested using different donwsamplings of the original database, representing different sensor positions,  and also the effect of scaling the original database have been studied. The results show that method presented is not only capable to reconstruct the original flow field, as in the case of the ABL, but also is capable to predict the temporal evolution of new reconstructed databases (also coming from sensor measurements), with errors smaller than $5-10$\%, depending on the type of downsampling and complexity of the case. The model has also been proved to perfectly work with noisy databases and also when the sensors are not collected equidistant in space, but grouped into some specific areas of the field.


The results obtained shows that the novel method, which is based on a very simple neural network architecture (minimizing computational costs and increasing generalization capabilities) is robust and supports all types of databases tested.


It has been shown that the neural network, in addition to reconstructing, is also able to predict in time. This, together with the low computational cost that has been obtained, is a very valuable tool that could save a lot of data simulation costs. 

Finally, summary tables with the RRMSE and MSE errors presented throughout the results analysis have been included in order to have an overview of the error order achieved with this neural network. The errors presented in the tables present the same order than the errors from simulations found in the literature of similar problems \cite{abadia2022predictive} \cite{le2017higher} \cite{kou2018reduced} \cite{le2018reduced}.
The present work achieves errors of the same order of magnitude in much less time, demonstrating its suitability for the reconstruction of fluid dynamics databases.

To sum up, this work proposes a hybrid physics-based machine learning model with a simple, robust and generalizable architecture, which allows reconstructing databases from very few points mimicking a sensors network and with a low computational cost. 

\begin{table}[H]
     \caption{Summary of the RRMSE error rate in reconstructing the different databases and computational cost.}
        \centering
        \resizebox{15cm}{!} {
        \begin{tabular}{ c | c | c | r | r | r | r | r}
        \toprule[0.8mm]
        \textbf{Database} & \multicolumn{2}{m{3.8cm}|}{\textbf{Method}} & \textbf{Variable 1} & \textbf{Variable 2} & \textbf{Variable 3} & \textbf{TOTAL}  & \textbf{Cost (s)} \\
        \midrule \midrule
        \multirow{4}{4cm}{\centering \textbf{ABL \\ 1:50}}  & \multirow{2}{3cm}{\centering NN Test 1} & No scaling & 0.12 & 0.00023 &0.00081 & 0.00023 & 93.9
         \tabularnewline \cline{3-8} & & Scaling & 0.12 & 0.00039 & 0.00056 & 0.00056 & 80.1
         \tabularnewline \cline{2-8} & \multirow{2}{3cm}{\centering NN Test 2} & No scaling & 0.73 & 0.00023 & 0.00081 & 0.00026 & 99.3
        \tabularnewline \cline{3-8} & & Scaling & 0.11& 0.00095 & 0.00044 & 0.00095 & 99.1\\
        
        \midrule \midrule
        \multirow{4}{4cm}{\centering \textbf{Cylinder 2D \\ 1:50}}  & \multirow{2}{3cm}{\centering NN} & No scaling & 14.1 & 29.0 & & 21.5 & 54.4
         \tabularnewline \cline{3-8} & & Scaling & 7.9& 22.4 & & 15.1 & 55.4
        \tabularnewline \cline{2-8} & \multirow{2}{3cm}{\centering NN (noise test)} & 5\% interference & 17.2 & 46.9 & & 20.4 & 20.4
        \tabularnewline \cline{3-8} & & 10\% interference & 18.1& 51.0 & & 34.5 & 22.6 \\
         
        \midrule \midrule
        \multirow{4}{4cm}{\centering \textbf{Cylinder 3D \\ 1:20}}  & \multirow{2}{3cm}{\centering NN Test 1} & No scaling & 11.3 & 28.5 & 89.3 & 14.4 & 169.5
        \tabularnewline \cline{3-8} & & Scaling & 10.1 & 27.9  & 90.7 & 12.3 & 300.8
        \tabularnewline \cline{2-8} & \multirow{2}{3cm}{\centering NN Test 2} & No scaling & 10.3 & 26.4 & 85.6 & 13.3 &123.8
         \tabularnewline \cline{3-8} & & Scaling & 9.5 & 29.9 & 86.3 & 13.6 & 196.6 \\
       \bottomrule[0.8mm]
       \end{tabular}
       }
     \label{tab:comparacionRRMSE}
\end{table}

\begin{table}[H]
     \caption{Summary of the MSE error made when reconstructing the different databases.}
        \centering
        \resizebox{15cm}{!} {
        \begin{tabular}{ c | c | c | r | r | r | r }
        \toprule[0.8mm]
        \textbf{Database} & \multicolumn{2}{m{3.8cm}|}{\textbf{Method}} & \textbf{Variable 1} & \textbf{Variable 2} & \textbf{Variable 3} & \textbf{TOTAL} \\
        \midrule \midrule
        \multirow{4}{4cm}{\centering \textbf{ABL \\ 1:50}}  & \multirow{2}{3cm}{\centering NN Test 1} & No scaling &0.00034 &0.052 &0.000006 &0.017
         \tabularnewline \cline{3-7} & & Scaling &0.0003 &0.30 &0.0000005 &0.10
         \tabularnewline \cline{2-7} & \multirow{2}{3cm}{\centering NN Test 2} & No scaling & 0.012& 0.052&0.000006 &0.021
        \tabularnewline \cline{3-7} & & Scaling & 0.0003&0.88 &0.000002  &0.29 \\
        
        \midrule \midrule
        \multirow{4}{4cm}{\centering \textbf{Cylinder 2D \\ 1:50}}  & \multirow{2}{3cm}{\centering NN} & No scaling & 0.016& 0.0057& &0.011
         \tabularnewline \cline{3-7} & & Scaling &0.0050 &0.0034 & &0.0042
         \tabularnewline \cline{2-7} & \multirow{2}{3cm}{\centering NN (noise test)} & 5\% interference & 0.025 & 0.0015 &  & 0.013
         \tabularnewline \cline{3-7} & & 10\% interference & 0.030 & 0.018 & & 0.024 \\
         
        \midrule \midrule
        \multirow{4}{4cm}{\centering \textbf{Cylinder 3D \\ 1:20}}  & \multirow{2}{3cm}{\centering NN Test 1} & No scaling &0.011 &0.0091 &0.00050  &0.0070
        \tabularnewline \cline{3-7} & & Scaling & 0.0091 & 0.0087 &0.00052 &0.0061
        \tabularnewline \cline{2-7} & \multirow{2}{3cm}{\centering NN Test 2} & No scaling &0.0095 &0.0079 &0.00046 &0.0059 
         \tabularnewline \cline{3-7} & & Scaling &0.0082  & 0.010&  0.00047&0.0063  \\
         
       \bottomrule[0.8mm]
       \end{tabular}
       }
     \label{tab:comparacionMSE}
\end{table}

\section*{Acknowledgements}
A.C. and S.L.C. acknowledge the grant PID2020-114173RB-I00 funded by MCIN/AEI/ 10.13039/501100011033. P.D, A.C. and S.L.C. acknowledge the support of Comunidad de Madrid through the call Research Grants for Young Investigators from Universidad Politécnica de Madrid. A.C. also acknowledges the support of Universidad Politécnica de Madrid, under the program ‘Programa Propio’.

\bibliography{sample}


\end{document}